\documentclass[12pt]{article}

\usepackage{amsfonts}
\usepackage{amsmath}
\usepackage{amssymb}
\usepackage{cite}
\usepackage{graphicx}
\usepackage[svgnames]{xcolor}			
\usepackage{ytableau,tikz}					
\usepackage[colorlinks=true,linkcolor=black,citecolor=black,urlcolor=black]{hyperref}                 		

\usepackage{ascmac} 

\usetikzlibrary{intersections, calc,shapes.geometric,patterns} 	

\newcommand{\ba}{\begin{eqnarray}}
\newcommand{\ea}{\end{eqnarray}}
\newcommand{\nn}{\nonumber}

\newcommand{\fr}[2]{\frac{#1}{#2}}

\newcommand{\tth}{q^h}
\newcommand{\ttv}{q^v}
\newcommand{\br}{{\bar{r}}}
\newcommand{\bs}{{\bar{s}}}
\newcommand{\baf}{{\bar{f}}}
\newcommand{\bA}{{\bar{A}}}
\newcommand{\bB}{{\bar{B}}}

\newcommand{\bk}{\bar{l}}

\allowdisplaybreaks[1]

\begin{document}

\begin{titlepage}

\begin{flushright}
UT-14-44
\end{flushright}

\vskip 12mm

\begin{center}
{\Large Recursive Method for Nekrasov partition function}\\
{\Large for classical Lie groups}
\vskip 2cm
{\Large Satoshi Nakamura, Futoshi Okazawa and Yutaka Matsuo}
\vskip 2cm
{\it Department of Physics, The University of Tokyo}\\
{\it Hongo 7-3-1, Bunkyo-ku, Tokyo 113-0033, Japan}
\end{center}
\vfill
\begin{abstract}
Nekrasov partition function
for the supersymmetric gauge theories with general Lie groups is not so far known 
in a closed form while there is a
definition in terms of the integral.
In this paper, as an intermediate step to derive
it, we give a recursion 
formula among partition functions, which can be derived from
the integral.  We apply the method to a toy model
which reflects the basic structure of partition functions
for BCD type Lie groups and obtained a closed expression
for the factor
associated with the generalized Young diagram.
\end{abstract}
\vfill
\end{titlepage}

\setcounter{footnote}{0}


\section{Introduction}
Starting from a seminal work by Seiberg and Witten \cite{Seiberg:1994rs},
the connection between supersymmetric gauge theories in 4D and
2D conformal field theories is getting more and more tight.
The first key step in such direction is the explicit derivation of
instanton partition function by Nekrasov and others
in the omega background \cite{Nekrasov:2003af,Nekrasov2003}.
They were written
as a sum over factors associated with (a set of) Young diagrams
which implied the connection with 2D theories.
Such correspondence becomes more explicit by the AGT conjecture
\cite{Alday:2009aq}
where the partition function
is directly related to the correlation function of Liouville theory.
This conjecture, originally made for $SU(2)$ gauge theories,
was later generalized to $SU(N)$ gauge groups
and has been proved for various set-ups in the last few years
\cite{Fateev:2009aw, Alba:2010qc, schiffmann2013cherednik, 
maulik2012quantum, Kanno:2013aha, Morozov2014, Matsuo:2014rba}. 
The application of Nekrasov formula is not limited to AGT conjecture,
it has been used 
to study the nonperturbative aspects of gauge theories, 
like dualities and global symmetry enhancements.

So it will be valuable 
to understand the structure of these partition functions and their exact form.
A natural next step in such development is to consider the general Lie groups. 
In this respect, however, 
while many efforts have done \cite{Nekrasov:2004vw, Marino:2004cn,Hollands:2010xa,Keller:2011ek,Fucito:2004gi},
the analysis was limited to the cases for small number
of instantons.  The origin of such difficulty is that we do not know how
to perform the integration that defines the partition function.
For the $SU(N)$ case, the integration reduces to the evaluation
of residues where poles are located in the shape of Young diagrams.
There is another way to evaluate this partition functions by use of the 
Hilbert scheme, which give the same result\cite{Nakajima:2003pg,nakajimalectures}.
On the other hand, for $Sp(N)$ and $SO(N)$ cases, similar poles
show up in the form of  ``the generalized Young diagrams" and the
algorithm to derive the residue in a closed form was not known.
There are also some difficulties which does not exist for $SU(N)$ case.
One is the appearance of double poles and the other is
the multiplicity factor associated with each generalized Young diagram\cite{Hollands:2010xa}.
Each of them is a tough question, which may not be solved at once.

In this paper, we give a partial answer to these problems.  We develop a
method to derive the explicit form of the residue
for the any generalized Young diagram.
It is a generalization of the recursion formula derived in \cite{Kanno:2013aha}.
For $SU(N)$ case, it 
represents an infinite dimensional nonlinear symmetry \cite{schiffmann2013cherednik}
which is equivalent to $W$ algebra.
We derive the recursive relation in a
similar form and give the explicit form of the residue as a product of factors
for generalized Young diagrams.  For the derivation of the formula, however,
we will focus on a ``simplified version" of integral
in order to avoid the complication coming from the double pole.

We organize the paper as follows.  In section 2, we review the Nekrasov partition function
for classical Lie groups in the form of the integration.  For the $SU(N)$ case, such
integration is explicitly performed.
We explain the derivation of recursion formula for $SU(N)$ case in two ways,
a derivation from product formula and another one from the integration.
In section 3, we apply the latter idea to derive a similar recursion formula
for the simplified model in a form that is readily generalized to other Lie groups.
In section 4, we solve the recursion formula and derive the partition function
as a product of factors for the generalized Young diagram.
In the discussion, we give some arguments how to fight with
other problems by using our strategy.

\section{Recursive construction of the Nekrasov partition function: $A_n$ case}
In 2002, Nekrasov introduced a way to calculate the non-perturbative corrections 
of partition functions for 4-dimensional $\mathcal{N}=2$ supersymmetric 
$SU(N)$ gauge theories\cite{Nekrasov:2003af}. 
Soon later, by the parameterizations of instanton moduli spaces by ADHM matrices,
he and Shadchin managed to generalize the technique to the cases 
when the gauge groups are classical and 
gave integral representations of the corrections\cite{Nekrasov:2004vw,Shadchin:2005mx}.

When the gauge group is unitary, one can perform the integrations exactly 
and these results are  heavily used especially for deeper understandings 
of AGT correspondences\cite{Alday:2009aq}. 

So we want to find such parallel formulae for other gauge groups. 
In this section, as a warm-up,
we explain how such  recursion formulae show up in the case of
the unitary 4-dimensional $\mathcal{N}=2$ super-Yang-Mills theory.
The idea is to perform the integration inductively.
We will see later that it can ben readily generalized to
the cases of other Lie groups.

\subsection{Integral representations of instanton partition function for $ABCD$}
Let $N$ and $k$ be the positive integer.  
Let $Z_{k}^{U(N)}$ be the $k$-instanton correction to the partition function 
of the $4$d $\mathcal{N}=2$ supersymmetric $U(N)$ Yang-Mills theory. 
In the following, we restrict ourselves to $N_f=0$ case for simplicity.
When the gauge coupling is small, this value is approximately a volume of the classical moduli space of $k$-instanton. 
This moduli space can be expressed by the famous ADHM matrices\cite{Atiyah:1978ri}, 
constrained by the ADHM equation and the internal gauge symmetry rotating instantons. 
Turning on the small background vector field, called the $\Omega$-background, 
on the spacetime $\mathbb{R}^{4}$, localization technique makes the volume into the form of a summation 
over the fixed points in the moduli space with respect to the background. 
Thus $Z_{k}^{U(N)}$ can be represented by the following integral;
\begin{eqnarray}\label{zun}
Z_{k}^{U(N)}=\frac{1}{k!}\left(\frac{\varepsilon}{2\pi i\varepsilon_{1}\varepsilon_{2}}\right)^{k}\int_{\mathbb{R}}\frac{d\phi_{1}\cdots d\phi_{k}}{\prod_{j=1}^{k}P(\phi_{j}-\varepsilon_{+})P(\phi_{j}+\varepsilon_{+})}\frac{\Delta(0)\Delta(\varepsilon)}{\Delta(\varepsilon_{1})\Delta(\varepsilon_{2})}
\end{eqnarray}
with the appropriate coefficient of the dynamical scale of the theory, and
\begin{eqnarray}
&P(x)=\prod_{l=1}^{N}(x-a_{l})\\
&\Delta(x)=\prod_{i<j}((\phi_{i}-\phi_{j})^{2}-x^{2}).
\end{eqnarray}
The $a_{l}$'s which appear in $P(x)$ mean the vev of the adjoint matter and are
real numbers.\footnote{The reality of $a_{l}$'s is just for the correct contour of the integration. 
After the integration, the value is analytically continued.}
The $\varepsilon_{1,2}$ parametrize the $\Omega$-background 
$(e^{i\varepsilon_{1}},e^{i\varepsilon_{2}})\in U(1)^{2}_{\Omega}$.
In the integration, we treat them to have positive imaginary parts. 
String theoretically, $a_{l}$'s represent the positions of D-branes and the $\Omega$ background comes 
from the Kalb-Ramond 2-form field $B$. 
We use  notations $\varepsilon:=\varepsilon_{1}+\varepsilon_{2}=:2\varepsilon_{+}$. 
The integration variables, $\phi$'s, are originally defined to 
parameterize the maximal torus of $\mathfrak{u}(k)$, the internal group rotating instantons. 
When we take the limit to 4d gauge theories, the integration range is replaced to those on the real axis.
The denominator of the integrand comes from the weights of the ADHM matrices under $U(k)\times U(N)\times U(1)^{2}_{\Omega}$, and the numerator comes from the constraint of the ADHM equation and of the internal gauge symmetry. This integral ends up with the summation over poles, where some of the weights of the ADHM matrices vanish. These correspond the fixed points of the moduli space
in the localization method.

The above arguments except for the last step are also applicable when one can coordinate similar classical moduli spaces. 
For classical groups, the ADHM construction gives such expressions. 
So there are similar integral representations of the instanton corrections for the other classical gauge groups. 
For super-Yang-Mills theories, the instanton partition functions are represented as follows;
\ba
Z_{k}^{SO(N=2n+\chi)}=\frac{(-1)^{k(N+1)}}{2^{k}k!}\left(\frac{\varepsilon}{2\pi i\varepsilon_{1}\varepsilon_{2}}\right)^{k}\int_{\mathbb{R}}\frac{d\phi_{1}\cdots d\phi_{k}}{\prod_{j=1}^{k}P(\phi_{j}-\varepsilon_{+})P(\phi_{j}+\varepsilon_{+})}\frac{\Delta(0)\Delta(\varepsilon)}{\Delta(\varepsilon_{1})\Delta(\varepsilon_{2})}\,,
\ea
\begin{eqnarray}
&P(x)=x^{\chi}\prod_{l=1}^{n}(x^{2}-a^{2}_{l})\,,\\
&\Delta(x)=\prod_{i<j}((\phi_{i}-\phi_{j})^{2}-x^{2})((\phi_{i}+\phi_{j})^{2}-x^{2})\,,
\end{eqnarray}
\begin{align}
Z_{k=2n+\chi}^{Sp(N)}=\frac{1}{2}\frac{(-1)^{n}}{2^{n-1}n!}&\left(\frac{\varepsilon}{2\pi i\varepsilon_{1}\varepsilon_{2}}\right)^{n}\left(-\frac{1}{2\varepsilon_{1}\varepsilon_{2}P(\varepsilon_{+})}\right)^{\chi}\nonumber\\
&\times\int_{\mathbb{R}}\frac{d\phi_{1}\cdots d\phi_{n}}{\prod_{j=1}^{n}P(\phi_{j}-\varepsilon_{+})P(\phi_{j}+\varepsilon_{+})(4\phi_{j}^{2}-\varepsilon_{1}^{2})(4\phi_{j}^{2}-\varepsilon_{2}^{2})}\frac{\Delta(0)\Delta(\varepsilon)}{\Delta(\varepsilon_{1})\Delta(\varepsilon_{2})}
\,,
\end{align}
\begin{eqnarray}
&P(x)=\prod_{l=1}^{N}(x^{2}-a_{l}^{2})\,,\\
&\Delta(x)=\left(\prod_{i<j}((\phi_{i}-\phi_{j})^{2}-x^{2})((\phi_{i}+\phi_{j})^{2}-x^{2})\right)\left(\prod_{j=1}^{n}(\phi_{j}^{2}-x^{2})\right)^{\chi}\,.
\end{eqnarray}
Again the integration variables $\phi$ represent the maximal torus of $\mathfrak{sp}(k)$ or $\mathfrak{so}(k)$ and
are replaced by the integration along $\mathbb{R}$ in the 4D limit. 
$\chi$ takes the value $0$ (resp.  $1$) when the instanton number $k$ is even (resp. odd).

While the integration formulae are similar to $U(N)$, there are some characteristic 
difference at the same time.
First, the factor $\Delta$ contains $\phi_i+\phi_j$ as well as the difference $\phi_i-\phi_j$.
Second,  the function $P$ contains the extra factors which gives rise to
extra poles near the origin $\phi_i=0$.  Third, the number of the integration variables
is $[k/2]$ instead of $k$ and we have extra factor which shows up 
when $k$ is odd.
We will discuss the implication of them later in the next section.

\subsection{Recursion relation for $U(N)$ cases}
The integration contours along the real axis in (\ref{zun}) can be closed
in the upper plane.  For $U(N)$ case, 
the integrand contains $P(\phi_j-\epsilon_+)$ and $\Delta(\epsilon_1)\Delta(\epsilon_2)$
in the denominator.  The first one gives rise to a pole at $\phi_j=a_l+\epsilon_+$
in the upper half plane.  The second one describe a pole in the neighbor of other integration variables,
$\phi_i-\phi_j=\epsilon_{1,2}$.  We pick these poles in each integration variables.
Some detail of the procedure will be explained later but for the moment
we accept that the poles are arranged in the shape of $N$ Young diagrams
$\overrightarrow{Y}=(Y^{(1)},Y^{(2)}\cdots,Y^{(N)})$ as
\begin{eqnarray}
a_{l}+\varepsilon_{+}+(n-1)\varepsilon_{1}+(m-1)\varepsilon_{2}
\end{eqnarray}
where $(n,m)$ are positive integers and describe a coordinate
on a Young diagram $Y^{(l)}$.  The instanton number $k$ is identical to
the number of boxes $k=\sum_{i}|Y^{(i)}|=k$.

$Z_{k}^{U(N)}$ is expressed as a sum:
\begin{eqnarray}
Z_{k}^{U(N)}=\sum_{\overrightarrow{Y}, |\overrightarrow{Y}|=k}Z_{k,\overrightarrow{Y}}^{U(N)}
\end{eqnarray}
where $Z_{k,\overrightarrow{Y}}^{U(N)}$ is the residues labeled by $\overrightarrow{Y}$.
The explicit formula\cite{Bruzzo:2002xf,Flume:2002az} is given by
\begin{align}\label{Zk}
Z_{k,\overrightarrow{Y}}^{U(N)}=\prod_{l,p=1}^{N}&\prod_{(i,j)\in Y^{(l)}}\frac{1}{a_{l}-a_{p}+(Y^{(l)}_{j}-i+1)\varepsilon_{1}+(j-(Y^{(p)T})_{i})\varepsilon_{2}}\nonumber\\
\times&\prod_{(i,j)\in Y^{(p)}}\frac{1}{a_{l}-a_{p}+(i-Y^{(p)}_{j})\varepsilon_{1}+((Y^{(l)T})_{i}-j+1)\varepsilon_{2}}
\end{align}
where $Y^{(p)T}$ is the transpose of  $Y^{(p)}$ and we denote the length of each row of a Young diagram by
$Y^{(l)}_j$.  They characterize the Young diagram:
\begin{eqnarray}
Y^{(l)}=(Y_{1}^{(l)}\ge Y_{2}^{(l)}\ge\cdots\ge Y_{h(l)}^{(l)}>0=Y_{h(l)+1}^{(l)}=Y_{h(l)+2}^{(l)}=\cdots)\,.
\end{eqnarray}
For example, a Young diagram\ytableausetup{smalltableaux,centertableaux}\ydiagram{3,2}\ytableausetup{nosmalltableaux,centertableaux}\ 
has the partition $Y_{1}=3, Y_{2}=2, Y_{\ge 3}=0\ $($h(1)=2$). 
Young diagrams appear in the formula through the ``arm" and ``leg" lengths,
which are given, for a Young diagram $Y$ and a position $s=(i,j)$
of the box (not necessarily in $Y$), by
\begin{eqnarray}
a_{Y}(s)=Y_{j}-i,\qquad
l_{Y}(s)=Y^{T}_{i}-j\,.
\end{eqnarray}
These integers can be represented diagrammatically; when $Y$ is of the form  
\begin{eqnarray}
Y=\ytableausetup{centertableaux,boxsize=1em}\begin{ytableau}		
\ & \ & \ &\ &\ &\ \\
\ & \ & s & \bullet & \bullet & \bullet \\
\ & \ & \circ & \ \\
\ & \ & \circ
\end{ytableau}\ytableausetup{nocentertableaux,boxsize=normal}
\ \ (s=(i,j)=(3,2))
\end{eqnarray}
then $a_{Y}(s)$ is equal to the number of black circles and $l_{Y}(s)$ is equal to the number of white circles (with appropriate signs).

In the following, we explain the recursive structure derived from (\ref{Zk}).
Such formula was derived in \cite{Kanno:2013aha} to show the
conformal Ward identity.  Here we provide a simplified analysis.
For the simplicity of argument, we describe
$N=1$ case, where the formula becomes
\begin{eqnarray}
Z_{k,Y}^{U(1)}=\prod_{s\in Y}\frac{1}{(a(s)+1)\varepsilon_{1}-l(s)\varepsilon_{2}}\frac{1}{-a(s)\varepsilon_{1}+(l(s)+1)\varepsilon_{2}}\,.
\end{eqnarray}
We compare the factor when we add a box to $Y$
 at the position $(\hat{n},\hat{m})$.  We write the new Young diagram as $Y_{+}$.
The addition of a box results in the change of factors at $i=\hat{n}$ or $j=\hat{m}$.
More explicitly, for each $s\in Y$,
\begin{eqnarray}
a_{Y_{+}}(s)=\begin{cases}a_{Y}(s)\ \ (j\neq\hat{m})\\
a_{Y}(s)+1=Y_{\hat{m}}-i+1=\hat{n}-i\ \ (j=\hat{m})
\end{cases}\\
l_{Y_{+}}(s)=\begin{cases}l_{Y}(s)\ \ (i\neq\hat{n})\\
l_{Y}(s)+1=Y^{T}_{\hat{n}}-j+1=\hat{m}-j\ \ (i=\hat{n}).
\end{cases}
\end{eqnarray}
Also, $a_{Y_{+}}(\hat{n},\hat{m})=l_{Y_{+}}(s)(\hat{n},\hat{m})=0$. The ratio between $Z_{k,Y}^{U(1)}$ and $Z_{k+1,Y_{+}}^{U(1)}$ becomes
\begin{align}
\frac{Z_{k+1,Y_{+}}^{U(1)}}{Z_{k,Y}^{U(1)}}&=\frac{1}{\varepsilon_{1}\varepsilon_{2}}\prod_{j=1}^{\hat{m}-1}
\frac{(\hat{n}-i)\varepsilon_{1}-(Y_{i}^{T}-\hat{m})\varepsilon_{2}}{(\hat{n}-i+1)\varepsilon_{1}-(Y_{i}^{T}-\hat{m})\varepsilon_{2}}\cdot\frac{(i+1-\hat{n})\varepsilon_{1}+(Y_{i}^{T}-\hat{m}+1)\varepsilon_{2}}{(i-\hat{n})\varepsilon_{1}+(Y_{i}^{T}-\hat{m}+1)\varepsilon_{2}}\nonumber\\
&\times\prod_{j=1}^{\hat{m}-1}\frac{(Y_{j}-\hat{n}+1)\varepsilon_{1}-(\hat{m}-j-1)\varepsilon_{2}}{(Y_{j}-\hat{n}+1)\varepsilon_{1}-(\hat{m}-j)\varepsilon_{2}}\cdot\frac{(\hat{n}-Y_{j})\varepsilon_{1}+(\hat{m}-j)\varepsilon_{2}}{(\hat{n}-Y_{j})\varepsilon_{1}+(\hat{m}-j+1)\varepsilon_{2}}.
\end{align}

A characteristic feature of the formula is that it is a product of factors
of the form $(x-\hat n)\epsilon_1+(y-\hat m)\epsilon_2$.  One may represent them
graphically by indicating the position $(x,y)$ on $Y$.
We assign  the factors  
in the denominator (resp. numerator) to the boxes with $-$ (resp. $+$) sign.
Interestingly these factors are located on the boundary of Young diagram $Y$
and there are some cancellation of factors in the numerator and the denominator.
Graphically the cancellation can be illustrated as:
\begin{align}
\frac{Z_{k+1,Y_{+}}^{U(1)}}{Z_{k,Y}^{U(1)}}&=\frac{1}{\varepsilon_{1}\varepsilon_{2}}\times
\begin{tikzpicture}[scale=.4,>=stealth,baseline=(current bounding box.center)]
\coordinate (A)at(0,0);
\coordinate (B)at(0,-10);
\coordinate (C)at(5,-10);
\coordinate (D)at(5,-7);
\coordinate (E)at(10,-7);
\coordinate (F)at(10,-3);
\coordinate (G)at(15,-3);
\coordinate (H)at(15,0);
\draw (A)--(B)--(C)--(D)--(E)--(F)--(G)--(H)--(A);
\draw [pattern=north west lines] (D) rectangle ([xshift=1cm, yshift=-1cm]D);
\draw [] ([xshift=2cm]B) rectangle ([xshift=3cm, yshift=1cm]B) node[pos=.5]{$+$};
\draw [] ([xshift=3cm]B) rectangle ([xshift=4cm, yshift=-1cm]B) node[pos=.5]{$+$};
\draw [] ([xshift=2cm]B) rectangle ([xshift=3cm, yshift=-1cm]B) node[pos=.5]{$-$};
\draw [] ([xshift=2cm]B) rectangle ([xshift=1cm, yshift=1cm]B) node[pos=.5]{$-$};
\draw [] ([yshift=1cm]E) rectangle ([xshift=1cm, yshift=2cm]E) node[pos=.5]{$+$};
\draw [] ([yshift=2cm]E) rectangle ([xshift=-1cm, yshift=3cm]E) node[pos=.5]{$+$};
\draw [] ([yshift=2cm]E) rectangle ([xshift=1cm, yshift=3cm]E) node[pos=.5]{$-$};
\draw [] ([yshift=3cm]E) rectangle ([xshift=-1cm, yshift=4cm]E) node[pos=.5]{$-$};
\draw[->,bend left,distance=0.7cm,ultra thick] ([xshift=3cm, yshift=-2cm]D) node[right]{$(\hat{n},\hat{m})$} to ([xshift=.5cm, yshift=-.5cm]D);
\draw[->,bend left,distance=0.7cm,ultra thick] ([xshift=2.5cm, yshift=3cm]B) node[above]{$(i,Y_{i}^{T})$} to ([xshift=2.75cm, yshift=0.75cm]B);
\draw[->,bend right,distance=0.7cm,ultra thick] ([xshift=-3cm, yshift=2.5cm]E) node[left]{$(Y_{j},j)$} to ([xshift=-.75cm, yshift=2.25cm]E);
\draw[dashed] (B) -- ([yshift=-2cm]B);
\draw[dashed] (C) -- ([yshift=-2cm]C);
\draw[dashed] (E) -- ([xshift=6cm]E);
\draw[dashed] (H) -- ([xshift=1cm]H);
\draw[<->,ultra thick]([yshift=-2cm]B)--([yshift=-2cm]C)node[right]{$i=1,\cdots,\hat{n}-1$};
\draw[<->,ultra thick]([xshift=6cm]E)--node[right]{$j=1,\cdots,\hat{m}-1$}([xshift=1cm]H);
\end{tikzpicture}\\
&=\frac{1}{\varepsilon_{1}\varepsilon_{2}}\times
\begin{tikzpicture}[scale=.4,baseline=(current bounding box.center)]
\coordinate (A)at(0,0);
\coordinate (B)at(0,-10);
\coordinate (C)at(5,-10);
\coordinate (D)at(5,-7);
\coordinate (E)at(10,-7);
\coordinate (F)at(10,-3);
\coordinate (G)at(15,-3);
\coordinate (H)at(15,0);
\draw (A)--(B)--(C)--(D)--(E)--(F)--(G)--(H)--(A);
\draw [pattern=north west lines] (D) rectangle ([xshift=1cm, yshift=-1cm]D);
\draw [] (B) rectangle ([xshift=1cm, yshift=-1cm]B) node[pos=.5]{$-$};
\draw [] (B) rectangle ([xshift=-1cm, yshift=1cm]B) node[pos=.5]{$-$};
\draw [] (C) rectangle ([xshift=1cm, yshift=-1cm]C) node[pos=.5]{$+$};
\draw [] (C) rectangle ([xshift=-1cm, yshift=1cm]C) node[pos=.5]{$+$};
\draw [] (E) rectangle ([xshift=1cm, yshift=-1cm]E) node[pos=.5]{$+$};
\draw [] (E) rectangle ([xshift=-1cm, yshift=1cm]E) node[pos=.5]{$+$};
\draw [] (F) rectangle ([xshift=1cm, yshift=-1cm]F) node[pos=.5]{$-$};
\draw [] (F) rectangle ([xshift=-1cm, yshift=1cm]F) node[pos=.5]{$-$};
\draw [] (G) rectangle ([xshift=1cm, yshift=-1cm]G) node[pos=.5]{$+$};
\draw [] (G) rectangle ([xshift=-1cm, yshift=1cm]G) node[pos=.5]{$+$};
\draw [] (H) rectangle ([xshift=1cm, yshift=-1cm]H) node[pos=.5]{$-$};
\draw [] (H) rectangle ([xshift=-1cm, yshift=1cm]H) node[pos=.5]{$-$};
\end{tikzpicture}.
\label{PAE}
\end{align}
After the cancellation, the residue is written as the product of the factors
associated with the corners of the Young diagram.
In order to represent the edges, we divide the Young diagram
by rectangles described by a set of integers $r_0:=0<r_1<r_2<\cdots<r_f$
and $s_1>\cdots>s_f>s_{f+1}:=0$ (see the figure below for $f=3$ case).
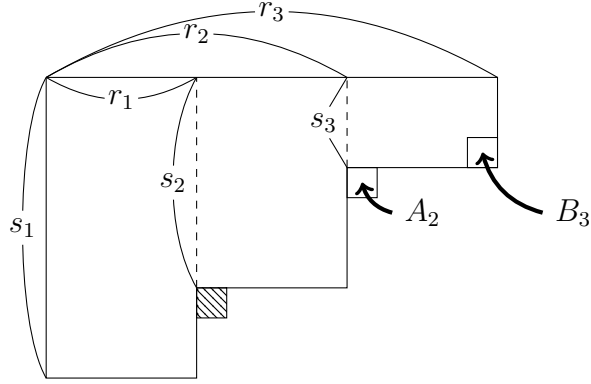
\begin{eqnarray}
\begin{tikzpicture}[scale=.4]
\coordinate (A)at(0,0);
\coordinate (B)at(0,-10);
\coordinate (C)at(5,-10);
\coordinate (D)at(5,-7);
\coordinate (E)at(10,-7);
\coordinate (F)at(10,-3);
\coordinate (G)at(15,-3);
\coordinate (H)at(15,0);
\draw (A)--(B)--(C)--(D)--(E)--(F)--(G)--(H)--(A);
\draw[dashed] (D)--([yshift=7cm]D);
\draw[dashed] (F)--([yshift=3cm]F);
\draw[bend right, distance=2.1cm] (A) to node [fill=white, inner sep=0.2pt,circle] {$s_{1}$} (B);
\draw[bend right, distance=2.1cm] ([yshift=7cm]D) to node [fill=white, inner sep=0.2pt,circle] {$s_{2}$} (D);
\draw[bend right, distance=2.1cm] ([yshift=3cm]F) to node [fill=white, inner sep=0.2pt,circle] {$s_{3}$} (F);
\draw[bend right] (A) to node [fill=white, inner sep=0.2pt,circle] {$r_{1}$} ([yshift=7cm]D);
\draw[bend left] (A) to node [fill=white, inner sep=0.2pt,circle] {$r_{2}$} ([yshift=3cm]F);
\draw[bend left] (A) to node [fill=white, inner sep=0.2pt,circle] {$r_{3}$} (H);
\draw [pattern=north west lines] (D) rectangle ([xshift=1cm, yshift=-1cm]D);
\draw [] (G) rectangle ([xshift=-1cm, yshift=1cm]G);
\draw [] (F) rectangle ([xshift=1cm, yshift=-1cm]F);
\draw[->,bend left,ultra thick] ([xshift=1.5cm,yshift=-1.5cm]F)node[right]{$A_{2}$}to([xshift=.5cm,yshift=-.5cm]F);
\draw[->,bend left,ultra thick] ([xshift=1.5cm,yshift=-1.5cm]G)node[right]{$B_{3}$}to([xshift=-.5cm,yshift=.5cm]G);
\end{tikzpicture}\nonumber
\end{eqnarray}
We define the factor which describe the edges:
\begin{eqnarray}
\begin{cases}
A_{i}=(r_{i}+1)\varepsilon_{1}+(s_{i+1}+1)\varepsilon_{2}+(a_{1}+\varepsilon_{+})\ \ (i=0,1,\cdots, f, r_{0}=0, s_{f+1}=0)\\
B_{i}=r_{i}\varepsilon_{1}+s_{i}\varepsilon_{2}+(a_{1}+\varepsilon_{+})\ \ (i=1,2,\cdots f).
\end{cases}
\end{eqnarray}
Then the diagrammatic formula is written as
\begin{eqnarray}
\frac{Z_{k+1,Y_{+}}^{U(1)}}{Z_{k,Y}^{U(1)}}=\frac{1}{\varepsilon_{1}\varepsilon_{2}}\frac{\prod_{j=1}^{f}\left(B_{j}-A_{l}+\varepsilon\right)\left(B_{j}-A_{l}\right)}{\prod_{i=0, i\neq l}^{f}\left(A_{i}-A_{l}\right)\left(A_{i}-\varepsilon-A_{l}\right)}
\end{eqnarray}
where $A_{l}=\begin{tikzpicture}[scale=.4,baseline=-8]\draw [pattern=north west lines] (0,0) rectangle (-1,-1);\end{tikzpicture}$.

The generalization for higher $N$ is straightforward.  Again we have similar cancellation of
factors along the boundaries on $Y^{(p)}$.
The explicit formula can be found in the appendix of \cite{Kanno:2013aha}.

\subsection{Recursion relation from integral formula}
For general Lie groups, the explicit form of partition function
that corresponds to (\ref{Zk}) is not known.
We use a recursion formula to derive them.
For that purpose we need to derive it directly from the integration 
such as (\ref{zun}).
Here we explain the procedure for $U(N)$ case.
We assign numbers to boxes in $\overrightarrow{Y}$ like
\begin{eqnarray}
\overrightarrow{Y}=\left(\ytableausetup{centertableaux,boxsize=1em}\ydiagram{3,2},\ydiagram{2,1},\cdots,\ydiagram{1}\right)\mapsto \left(\begin{ytableau}1&2&3\\4&5\end{ytableau},\begin{ytableau}6&7\\8\end{ytableau},\cdots,\begin{ytableau}k\end{ytableau}\ytableausetup{nocentertableaux,boxsize=normal}\right)\,.
\end{eqnarray}
Namely $i\in Y^{(p)}$ and $j\in Y^{(q)}$ should satisfy $i<j$ if $p<q$.  
Similarly for the boxes in the same Young diagram,
we assign numbers from top to bottom and left to right as shown in the diagrams.

The poles of the integration (\ref{zun}) are described by $\hat{\phi}_{i} (i=1,\cdots k)$
\begin{eqnarray}
\hat{\phi}_{i}=a_{l}+\varepsilon_{+}+(n_{i}-1)\varepsilon_{1}+(m_{i}-1)\varepsilon_{2}
\end{eqnarray}
where we assume the box $i$ is in $Y^{(l)}$ and 
$(n_{i},m_{i})$ is the position of the box in the diagram $Y^{(l)}$. 
For example, $l(4)=1$ and $(n_{4},m_{4})=(1,2)$. 
The residues at $\phi_i=\hat\phi_i$ becomes
\begin{align}
Z_{k,\overrightarrow{Y}}^{U(N)}=& \frac{n_{\vec Y}}{k!}
\left(\frac{\varepsilon}{\varepsilon_{1}\varepsilon_{2}}\right)^{k}\oint_{\hat{\phi}_{k}}\frac{d\phi_{k}}{2\pi i}\cdots \oint_{\hat{\phi}_{1}}\frac{d\phi_{1}}{2\pi i}\frac{1}{\prod_{j=1}^{k}P(\phi_{j}-\varepsilon_{+})P(\phi_{j}+\varepsilon_{+})}\frac{\Delta(0)\Delta(\varepsilon)}{\Delta(\varepsilon_{1})\Delta(\varepsilon_{2})}
\end{align}
where $\oint_{\hat{\phi}}$ indicates that this integration is done along a sufficiently small circle around $\hat{\phi}$.  
Here $n_{\vec Y}$ is the combinational factor.
In the case of $U(N)$, this factor is $k!$, 
which can be interpreted as coming from different choices of poles $\phi_i=\hat\phi_{\sigma(i)}$
with $\sigma\in \mathfrak{S}_{k}$ and cancels the factor in the denominator.
This combinatorial factor will be more complicated for other Lie groups.

Now we add one extra box to $\overrightarrow{Y}$ to get another vector 
$\overrightarrow{Y_{+}}$ of Young diagrams. 
We set the extra box at a position $(\hat n,\hat m)$ in $Y^{(l)}$. In the language of $\hat{\phi}$,
\begin{eqnarray}
\hat{\phi}_{k+1}=\hat{\phi}=a_{\hat{l}}+\varepsilon_{+}+(\hat{n}-1)\varepsilon_{1}+(\hat{m}-1)\varepsilon_{2}
\end{eqnarray}
We get
\begin{eqnarray}
Z_{k+1,\overrightarrow{Y_{+}}}^{U(N)}=\left(\frac{\varepsilon}{\varepsilon_{1}\varepsilon_{2}}\right)^{k+1}\oint_{\hat{\phi}_{k+1}}\frac{d\phi_{k+1}}{2\pi i}\oint_{\hat{\phi}_{k}}\frac{d\phi_{k}}{2\pi i}\cdots \oint_{\hat{\phi}_{1}}\frac{d\phi_{1}}{2\pi i}\frac{1}{\prod_{j=1}^{k+1}P(\phi_{j}-\varepsilon_{+})P(\phi_{j}+\varepsilon_{+})}\frac{\Delta(0)\Delta(\varepsilon)}{\Delta(\varepsilon_{1})\Delta(\varepsilon_{2})}\nonumber\\
\ 
\end{eqnarray}
After the integrations with respect to $\phi_{1},\cdots,\phi_{k}$, we get
\begin{eqnarray}
Z_{k+1,\overrightarrow{Y_{+}}}^{U(N)}=Z_{k,\overrightarrow{Y}}^{U(N)}\frac{\varepsilon}{\varepsilon_{1}\varepsilon_{2}}\oint_{\hat{\phi}}\frac{d\phi}{2\pi i}\frac{1}{P(\phi-\varepsilon_{+})P(\phi+\varepsilon_{+})}\prod_{j=1}^{k}\frac{(\phi-\hat{\phi}_{j})^{2}\left((\phi-\hat{\phi}_{j})^{2}-\varepsilon^{2}\right)}{\left((\phi-\hat{\phi}_{j})^{2}-\varepsilon_{1}^{2}\right)\left((\phi-\hat{\phi}_{j})^{2}-\varepsilon_{2}^{2}\right)}
\end{eqnarray}
Again we have a diagrammatic representation of the integrand.
We assign boxes with sign in the Young table the factor $\phi-\hat\phi_j$ in the numerator  (denominator).
This integrand is represented diagrammatically;
\begin{eqnarray}
\frac{(\phi-\hat{\phi}_{j})^{2}\left((\phi-\hat{\phi}_{j})^{2}-\varepsilon^{2}\right)}{\left((\phi-\hat{\phi}_{j})^{2}-\varepsilon_{1}^{2}\right)\left((\phi-\hat{\phi}_{j})^{2}-\varepsilon_{2}^{2}\right)}=
\begin{tikzpicture}[scale=.5,baseline=(current bounding box.center)]
\coordinate (A)at(0,0);
\coordinate (B)at(0,-10);
\coordinate (C)at(5,-10);
\coordinate (D)at(5,-7);
\coordinate (E)at(10,-7);
\coordinate (F)at(10,-3);
\coordinate (G)at(15,-3);
\coordinate (H)at(15,0);
\draw (A)--(B)--(C)--(D)--(E)--(F)--(G)--(H)--(A);
\coordinate (I) at (5,-3);
\draw [](I) rectangle ([xshift=1cm,yshift=-1cm]I) node[pos=.5]{$+2$};
\draw [](I) rectangle ([xshift=-1cm,yshift=1cm]I) node[pos=.5]{$+$};
\draw []([xshift=1cm,yshift=-1cm]I) rectangle ([xshift=2cm,yshift=-2cm]I) node[pos=.5]{$+$};
\draw [](I) rectangle ([xshift=1cm,yshift=1cm]I) node[pos=.5]{$-$};
\draw [](I) rectangle ([xshift=-1cm,yshift=-1cm]I) node[pos=.5]{$-$};
\draw []([xshift=1cm]I) rectangle ([xshift=2cm,yshift=-1cm]I) node[pos=.5]{$-$};
\draw []([yshift=-1cm]I) rectangle ([xshift=1cm,yshift=-2cm]I) node[pos=.5]{$-$};
\draw ([xshift=1cm]I) rectangle ([xshift=2cm,yshift=1cm]I) ;
\draw ([yshift=-1cm]I) rectangle ([xshift=-1cm,yshift=-2cm]I) ;
\draw [->,bend right,ultra thick] (2,-6) node[left]{$\hat{\phi}_{j}$} to ([xshift=.5cm,yshift=-.75cm]I);
\node at (11,-8) {$Y^{(l(i))}$};
\end{tikzpicture}
\end{eqnarray}
Then we have cancellation for factors assigned to boxes in the Young tables.
We arrive at the ratio of $Z$'s;
\begin{eqnarray}
\frac{Z_{k+1,\overrightarrow{Y_{+}}}^{U(N)}}{
Z_{k,\overrightarrow{Y}}^{U(N)}
}=\frac{\varepsilon}{\varepsilon_{1}\varepsilon_{2}}\oint_{\hat{\phi}}\frac{d\phi}{2\pi i}\prod_{l=1}^{N}
\begin{tikzpicture}[scale=.4,baseline=(current bounding box.center)]
\coordinate (A)at(0,0);
\coordinate (B)at(0,-10);
\coordinate (C)at(5,-10);
\coordinate (D)at(5,-7);
\coordinate (E)at(10,-7);
\coordinate (F)at(10,-3);
\coordinate (G)at(15,-3);
\coordinate (H)at(15,0);
\draw (A)--(B)--(C)--(D)--(E)--(F)--(G)--(H)--(A);
\draw [] (D) rectangle ([xshift=1cm, yshift=-1cm]D) node[pos=.5]{$-$};
\draw [] (D) rectangle ([xshift=-1cm, yshift=1cm]D) node[pos=.5]{$-$};
\draw [] (B) rectangle ([xshift=1cm, yshift=-1cm]B) node[pos=.5]{$-$};
\draw [] (B) rectangle ([xshift=-1cm, yshift=1cm]B) node[pos=.5]{$-$};
\draw [] (C) rectangle ([xshift=1cm, yshift=-1cm]C) node[pos=.5]{$+$};
\draw [] (C) rectangle ([xshift=-1cm, yshift=1cm]C) node[pos=.5]{$+$};
\draw [] (E) rectangle ([xshift=1cm, yshift=-1cm]E) node[pos=.5]{$+$};
\draw [] (E) rectangle ([xshift=-1cm, yshift=1cm]E) node[pos=.5]{$+$};
\draw [] (F) rectangle ([xshift=1cm, yshift=-1cm]F) node[pos=.5]{$-$};
\draw [] (F) rectangle ([xshift=-1cm, yshift=1cm]F) node[pos=.5]{$-$};
\draw [] (G) rectangle ([xshift=1cm, yshift=-1cm]G) node[pos=.5]{$+$};
\draw [] (G) rectangle ([xshift=-1cm, yshift=1cm]G) node[pos=.5]{$+$};
\draw [] (H) rectangle ([xshift=1cm, yshift=-1cm]H) node[pos=.5]{$-$};
\draw [] (H) rectangle ([xshift=-1cm, yshift=1cm]H) node[pos=.5]{$-$};
\node at (5,-4) {$Y^{(l)}$};
\end{tikzpicture}
\end{eqnarray}
We find that there are simple poles at boxes with $-$ sign in the Young diagram
as pairs, one outside of the edge and the other inside of the edge.
Again the cancellation mechanics of the multiplicity implies the
poles inside the Young diagram do not contribute.
Now we get sum of residues by picking a pole at 
$\phi=\hat{\phi}=a_{\hat{l}}+\varepsilon_{+}+(\hat{n}-1)\varepsilon_{1}+(\hat{m}-1)\varepsilon_{2}
$.  As the diagram suggests,  the residue is the same as (\ref{PAE}).
\section{Recursion relation for $BCD$}
\subsection{New features for $BCD$ and a simplified integral}
Now we focus on the $BCD$ cases. While the appearance is similar,
there are some new features as we mentioned in the previous section.
Every material was explained in detail in the appendix of \cite{Hollands:2010xa}.
Let us review them in more detail to make this paper self-contained.

\subsubsection*{Generalized Young diagram}
What is crucially different from the unitary case is the definitions of the $\Delta(x)$'s. 
These contain factors like $\phi_{i}+\phi_{j}-x$ which implies we may pick
a pole at $\phi_i=-\phi_j+\epsilon_{1,2}$.
We have to take care of not only the location of poles
$\{\hat \phi_{1},\cdots,\hat\phi_{n}\}$, but also their mirror image $\{-\hat \phi_{1},\cdots,-\hat\phi_{n}\}$.

The shapes of the poles are not necessarily (vectors of) Young diagrams.  
Regarding this point,  Hollands, Keller and Song proposed the notion of a ``generalized Young diagram"\cite{Hollands:2010xa}, 
which is defined as the subtraction $Y-Y'$ of two Young diagrams $Y,Y'$ ($Y$ contains $Y'$).
 In other words, a generalized Young diagram must slope from the upper right to the lower left. For example,

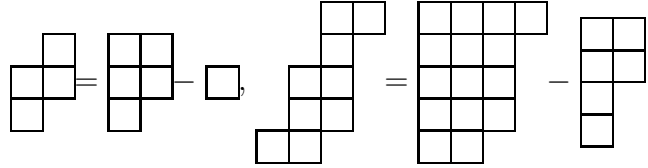
\begin{center}
\ytableausetup{centertableaux,boxsize=1em}
\ydiagram{1+1,2,1} $=$ \ydiagram{2,2,1} $-$ \ydiagram{1} ,\ 
\ydiagram{2+2,2+1,1+2,1+2,2} $=$ \ydiagram{4,3,3,3,2} $-$ \ydiagram{2,2,1,1}
\ytableausetup{nocentertableaux,boxsize=normal}
\end{center}
are generalized Young diagrams, but
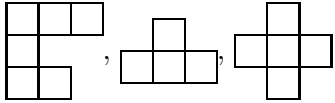
\begin{center}
\ytableausetup{centertableaux,boxsize=1em}
\ydiagram{3,1,2}  ,\  \ydiagram{1+1,3}  ,\  \ydiagram{1+1,3,1+1}
\ytableausetup{nocentertableaux,boxsize=normal}
\end{center}
are not. It is similar to the ``relative Young diagram" which 
appears in the literature.  It is also obtained
by subtracting a Young diagram from another one.
The difference between the two is that boxes in the generalized Young diagram
should be connected on the edges while the relative one may not be.

\subsubsection*{Location of Anchor and multiplicity of the diagram}
The other related difference is that the location of the ``anchor",
the location of the pole from $P(x)$.  For $U(N)$ case, such pole at
$\hat\phi=a_l+\epsilon_+$ is fixed at the top-left corner of the Young diagram.
On the other hand, in the generalized Young diagram one may in principle
put the anchor to any box in the diagram.

\ytableausetup{centertableaux,boxsize=1em}
For example, let us consider the $Sp(1)$ 4-instanton correction. 
There are two integration variable $\phi_{1,2}$ and two boxes.
One way of picking up poles is to take  $\phi_{1}=a_{1}+\varepsilon_{+}$ and $\phi_{2}=\phi_{1}+\varepsilon_{1}$.
In this case, we write the tableau as $\begin{ytableau}\times&\end{ytableau}_{a_{1}+\varepsilon_{+}}$,
where we indicate the location of the anchor by ``$\times$" and put their value as suffix.
We have other choice of picking poles, namely,  $\phi_{1}=-\phi_{2}+\varepsilon_{1}$ and $\phi_{2}=a_{1}+\varepsilon_{+}$.
This time, the diagram associated with it is written as, $\begin{ytableau}\ &\times\end{ytableau}_{a_{1}+\varepsilon_{+}}$.

The number of choice of poles for the same diagram, the multiplicity of the diagram, becomes more complicated.
For example, there are two other ways to obtain the diagram $\begin{ytableau}\times&\end{ytableau}_{a_{1}+\varepsilon_{+}}$,
namely $\phi_1=\phi_2+\epsilon_1, \phi_2=a_1+\epsilon_+$ and $\phi_1=-\phi_2+\epsilon_1,\phi_2=a_1+\epsilon_++\epsilon_1$.
So this diagram has multiplicity 3.  On the other hand, the diagram  $\begin{ytableau}\ &\times\end{ytableau}_{a_{1}+\varepsilon_{+}}$
has multiplicity 1.  Unlike the $U(N)$ case where the multiplicity for each Young diagram
is fixed to $|Y|!$, the multiplicity for generalized Young diagram depends on
the location of the anchor and the closed formula for them
is by itself a challenging problem.

\subsubsection*{Poles near the origin}
The next difference from the $U(N)$ cases is that there are poles near origin.
For $Sp(N)$ cases, for example,   in addition to the poles
at $\phi=\pm a_i$ we have extra poles at  $\phi=\pm \epsilon_i/2$.  
While the contributions from the first one can be interpreted
as the contribution of $D$-brane (and its mirror), the second one should be
interpreted as the contributions from the orientifold.  We have to combine the
contributions from both parts.
For example, the two instanton partition function 
$Z^{Sp(1)}_{2}$ is written in their expanded forms as
\ytableausetup{centertableaux,boxsize=1em}
\begin{eqnarray}
Z^{Sp(1)}_{2}=\left(\begin{ytableau}\times\end{ytableau}_{a_{1}+\varepsilon_{+}}+\begin{ytableau}\times\end{ytableau}_{-a_{1}+\varepsilon_{+}}\right)+\left(\begin{ytableau}\times\end{ytableau}_{\frac{\varepsilon_{1}}{2}}+\begin{ytableau}\times\end{ytableau}_{\frac{\varepsilon_{2}}{2}}\right)
\end{eqnarray} 
where $\begin{ytableau}\times\end{ytableau}_{x}$ means the residue at the pole where the box with $\times$ is at $x$. 
The first term comes from 2-instantons near D-branes and the second comes from ones near the orientifold plane.
Due to the isomorphism that $Sp(1)\simeq SU(2)$, the partition function should
be identical to that of $SU(2)$. In above example,
one may confirm that it is identical to $Z^{SU(2)}_{2}$ after some algebra.
Similarly the four-instanton partition function can be written graphically as,
\begin{align}
Z^{Sp(1)}_{4}=&\left(
3\times\begin{ytableau}\times\\ \ \end{ytableau}_{a_{1}+\varepsilon_{+}}
+3\times\begin{ytableau}\times\\ \ \end{ytableau}_{-a_{1}+\varepsilon_{+}}
+\begin{ytableau}\ \\ \times\end{ytableau}_{a_{1}+\varepsilon_{+}}
+\begin{ytableau}\ \\ \times\end{ytableau}_{-a_{1}+\varepsilon_{+}}
+3\times\begin{ytableau}\times&\end{ytableau}_{a_{1}+\varepsilon_{+}}
+3\times\begin{ytableau}\times&\end{ytableau}_{-a_{1}+\varepsilon_{+}}\right.\nonumber\\
&\left.+\begin{ytableau}\ &\times\end{ytableau}_{a_{1}+\varepsilon_{+}}
+\begin{ytableau}\ &\times\end{ytableau}_{-a_{1}+\varepsilon_{+}}\right)\nonumber\\
&+\left(3\times\begin{ytableau}\times\\ \ \end{ytableau}_{\frac{\varepsilon_{2}}{2}}+3\times\begin{ytableau}\times&\end{ytableau}_{\frac{\varepsilon_{1}}{2}}+\begin{ytableau}\ &\times\end{ytableau}_{\frac{\varepsilon_{2}}{2}}+\begin{ytableau}\ \\ \times\end{ytableau}_{\frac{\varepsilon_{1}}{2}}+2\times\begin{ytableau}\ \\ \times\end{ytableau}_{\varepsilon_{+}}+\begin{ytableau}\times\\ \ \end{ytableau}_{0}+\begin{ytableau}\times&\end{ytableau}_{0}+2\times\left(\begin{ytableau}\times\end{ytableau}_{\frac{\varepsilon_{1}}{2}},\begin{ytableau}\times\end{ytableau}_{\frac{\varepsilon_{2}}{2}}\right)\right)\nonumber\\
&+2\times\left(\left(\begin{ytableau}\times\end{ytableau}_{a_{1}+\varepsilon_{+}},\begin{ytableau}\times\end{ytableau}_{\frac{\varepsilon_{2}}{2}}\right)+\left(\begin{ytableau}\times\end{ytableau}_{-a_{1}+\varepsilon_{+}},\begin{ytableau}\times\end{ytableau}_{\frac{\varepsilon_{2}}{2}}\right)+\left(\begin{ytableau}\times\end{ytableau}_{a_{1}+\varepsilon_{+}},\begin{ytableau}\times\end{ytableau}_{\frac{\varepsilon_{1}}{2}}\right)+\left(\begin{ytableau}\times\end{ytableau}_{-a_{1}+\varepsilon_{+}},\begin{ytableau}\times\end{ytableau}_{\frac{\varepsilon_{1}}{2}}\right)\right)\nonumber\\(=&Z^{SU(2)}_{4})\,.
\end{align} 
The coefficients in front of the diagram imply the multiplicity.
The last term comes from the cross term of two 2-instantons.
\ytableausetup{nocentertableaux,boxsize=normal}
The contributions from poles take very different form compared with $SU(2)$ case.
After the summation, however, the two partition functions become identical
in a very nontrivial way.

Picking the poles near the origin needs the extra care since
the generalized Young diagram may involve extra anchors
since they are located near the origin.  It induces the second order
pole in the integration.  Also, it may be possible to have configurations
like closed cycles \cite{Hollands:2010xa} which contain no anchor
in the diagram.

\subsubsection*{Simplified model}
Having seen the extra complications to evaluate the integration,
it may be a good idea to focus on one of them and solve it
explicitly.  In this paper, we concentrate on the residue associated with
the generalized Young diagram with one anchor.
It consists of the evaluation of simple poles and the problem 
can be reduced to the recursion relations.
For that purpose, instead of the evaluation of BCD type integral directly,
we focus on a simplified version
\begin{eqnarray}
\tilde{Z}_{n}=\frac{1}{2}\frac{(-1)^{n}}{2^{n-1}n!}\left(\frac{\varepsilon}{2\pi i\varepsilon_{1}\varepsilon_{2}}\right)^{n}\int_{\mathbb{R}}\frac{d\phi_{1}\cdots d\phi_{n}}{\prod_{j=1}^{n}(\phi_{j}^{2}-b^{2})
}\frac{\Delta'(0)\Delta'(\varepsilon)}{\Delta'(\varepsilon_{1})\Delta'(\varepsilon_{2})}
\end{eqnarray}
where $b$ is a constant with a positive imaginary part and
\begin{eqnarray}
\Delta'(x)=\left(\prod_{i<j}((\phi_{i}-\phi_{j})^{2}-x^{2})((\phi_{i}+\phi_{j})^{2}-x^{2})\right).
\end{eqnarray}

Compared with the original integral, we simplified $P(x)$ to have a pole only at $b$.
In the meanwhile we mostly keep
 the form of $\Delta(x)$ to retain the structure of mirrors
in the analysis.
In the following of this paper, we concentrate on this simplified model. 
In this section, we derive recursion relations near D-branes (or $b$) and in the next section we solve them. 

\subsection{Recursion relation from integral formula}
In the following of this paper, we focus on the contributions
from a single D-brane.  In principle, one may combine such factors
with the contribution from orientifold plane to obtain
the general partition function.  For the simplicity of the computation, we use the expression
for the simplified model. We derive the recursion relations of fixed points near $\pm b$ in this model. 
In other words, we compare the integrals that are associated with two generalized Young diagrams.

Let $Y$ be a generalized Young diagram which is given by subtracting a Young diagram $Y_2$  from another $Y_{1}$, like
\begin{eqnarray}
\begin{tikzpicture}[scale=.3,baseline=(current bounding box.center)]
\coordinate (A)at(0,0);
\coordinate (B)at(0,-10);
\coordinate (C)at(5,-10);
\coordinate (D)at(5,-7);
\coordinate (E)at(10,-7);
\coordinate (F)at(10,-3);
\coordinate (G)at(15,-3);
\coordinate (H)at(15,0);
\coordinate (A')at(0,0);
\coordinate (B')at(0,-6);
\coordinate (C')at(3,-6);
\coordinate (D')at(3,-3);
\coordinate (E')at(8,-3);
\coordinate (F')at(8,0);
\draw (B')--(B)--(C)--(D)--(E)--(F)--(G)--(H)--(F')--(E')--(D')--(C')--(B') ;
\node at (7,-5) {$Y$};
\end{tikzpicture}
=\ 
\begin{tikzpicture}[scale=.3,baseline=(current bounding box.center)]
\coordinate (A)at(0,0);
\coordinate (B)at(0,-10);
\coordinate (C)at(5,-10);
\coordinate (D)at(5,-7);
\coordinate (E)at(10,-7);
\coordinate (F)at(10,-3);
\coordinate (G)at(15,-3);
\coordinate (H)at(15,0);
\draw (A)--(B)--(C)--(D)--(E)--(F)--(G)--(H)--(A) ;
\node at (7,-5) {$Y^{(1)}$};
\coordinate (A')at(0,0);
\coordinate (B')at(0,-6);
\coordinate (C')at(3,-6);
\coordinate (D')at(3,-3);
\coordinate (E')at(8,-3);
\coordinate (F')at(8,0);
\draw[dashed, ] (A')--(B')--(C')--(D')--(E')--(F')--(A');
\node at (2,-2) {$Y^{(2)}$};
\end{tikzpicture}
\end{eqnarray}
and $\{\hat{\phi}_{j} |j=1,\cdots,n\}$ be a set of some poles of the simplified integral, whose values and their minuses $\{\pm\hat{\phi}_{j}\}$ make the diagram $Y$ and its mirror diagram $-Y$. So we have
\ba
\frac{\tilde{Z}_{n,Y}}{n_{Y}} &=& \frac{(-1)^{n}}{2^{n}n!} \left( \frac{\varepsilon}{\varepsilon_1\varepsilon_2}\right)^{n}\oint_{\hat{\phi}_n} \frac{{\rm d}\phi_n}{2\pi i} \cdots \oint_{\hat{\phi}_1} \frac{{\rm d}\phi_1}{2\pi i} \frac{\Delta'(0)\Delta'(\varepsilon)}{\Delta'(\varepsilon_1)\Delta'(\varepsilon_2)} \prod_{j=1}^{n} \fr{1}{\phi_{j}^{2}-b^{2}}. \nn
\ea
where $n_Y$ is the combinational factor for $Y$ discussed in the last section. 

We add one extra box to $Y$ and label the 1-box-added diagram $Y_+$. 
We write the position of the extra box as
\ba
\hat{\phi}_{n+1} = \hat{\phi} = b+(\hat{n}-1)\varepsilon_1 + (\hat{m}-1)\varepsilon_2. \nn
\ea 
As in the previous section, we get the ratio of two integrals;
\ba
\frac{\tilde{Z}_{n+1,Y_+}}{n_{Y_+}}\left(\frac{\tilde{Z}_{n,Y}}{n_Y} \right)^{-1}\!\!\!\!\!\!\!\! &=& \frac{-1}{2(n+1)} \frac{\varepsilon}{\varepsilon_1\varepsilon_2} \oint_{\hat{\phi}} \frac{{\rm d}\phi}{2\pi i} \nn \\
& &\times \prod_{j=1}^n \frac{(\phi+\hat{\phi}_j)^2(\phi-\hat{\phi}_j)^2((\phi+\hat{\phi}_j)^2-\varepsilon^2)((\phi-\hat{\phi}_j)^2-\varepsilon^2)}{((\phi+\hat{\phi}_j)^2-\varepsilon_1^2)((\phi-\hat{\phi}_j)^2-\varepsilon_1^2)((\phi+\hat{\phi}_j)^2-\varepsilon_2^2)((\phi-\hat{\phi}_j)^2-\varepsilon_2^2)} \nn \\
& &\times \frac{1}{\phi^{2}-b^{2}}\,. \nn
\ea
This integrand can be expressed diagrammatically as a product of factors coming from $Y$ and $-Y$;
\begin{align}
&\prod_{j=1}^n \frac{(\phi+\hat{\phi}_j)^2(\phi-\hat{\phi}_j)^2((\phi+\hat{\phi}_j)^2-\varepsilon^2)((\phi-\hat{\phi}_j)^2-\varepsilon^2)}{((\phi+\hat{\phi}_j)^2-\varepsilon_1^2)((\phi-\hat{\phi}_j)^2-\varepsilon_1^2)((\phi+\hat{\phi}_j)^2-\varepsilon_2^2)((\phi-\hat{\phi}_j)^2-\varepsilon_2^2)}\nonumber\\
=&
\begin{tikzpicture}[scale=.4,baseline=(current bounding box.center)]
\coordinate (A)at(0,0);
\coordinate (B)at(0,-10);
\coordinate (C)at(5,-10);
\coordinate (D)at(5,-7);
\coordinate (E)at(10,-7);
\coordinate (F)at(10,-3);
\coordinate (G)at(15,-3);
\coordinate (H)at(15,0);
\coordinate (A')at(0,0);
\coordinate (B')at(0,-6);
\coordinate (C')at(3,-6);
\coordinate (D')at(3,-3);
\coordinate (E')at(8,-3);
\coordinate (F')at(8,0);
\draw (B')--(B)--(C)--(D)--(E)--(F)--(G)--(H)--(F')--(E')--(D')--(C')--(B') ;
\node at (7,-5) {$Y$};
\draw [] (D) rectangle ([xshift=1cm, yshift=-1cm]D) node[pos=.5]{$-$};
\draw [] (D) rectangle ([xshift=-1cm, yshift=1cm]D) node[pos=.5]{$-$};
\draw [] (B) rectangle ([xshift=1cm, yshift=-1cm]B) node[pos=.5]{$-$};
\draw [] (B) rectangle ([xshift=-1cm, yshift=1cm]B) node[pos=.5]{$-$};
\draw [] (C) rectangle ([xshift=1cm, yshift=-1cm]C) node[pos=.5]{$+$};
\draw [] (C) rectangle ([xshift=-1cm, yshift=1cm]C) node[pos=.5]{$+$};
\draw [] (E) rectangle ([xshift=1cm, yshift=-1cm]E) node[pos=.5]{$+$};
\draw [] (E) rectangle ([xshift=-1cm, yshift=1cm]E) node[pos=.5]{$+$};
\draw [] (F) rectangle ([xshift=1cm, yshift=-1cm]F) node[pos=.5]{$-$};
\draw [] (F) rectangle ([xshift=-1cm, yshift=1cm]F) node[pos=.5]{$-$};
\draw [] (G) rectangle ([xshift=1cm, yshift=-1cm]G) node[pos=.5]{$+$};
\draw [] (G) rectangle ([xshift=-1cm, yshift=1cm]G) node[pos=.5]{$+$};
\draw [] (H) rectangle ([xshift=1cm, yshift=-1cm]H) node[pos=.5]{$-$};
\draw [] (H) rectangle ([xshift=-1cm, yshift=1cm]H) node[pos=.5]{$-$};
\draw [] (C') rectangle ([xshift=1cm, yshift=-1cm]C') node[pos=.5]{$-$};
\draw [] (C') rectangle ([xshift=-1cm, yshift=1cm]C') node[pos=.5]{$-$};
\draw [] (E') rectangle ([xshift=1cm, yshift=-1cm]E') node[pos=.5]{$-$};
\draw [] (E') rectangle ([xshift=-1cm, yshift=1cm]E') node[pos=.5]{$-$};
\draw [] (B') rectangle ([xshift=1cm, yshift=-1cm]B') node[pos=.5]{$+$};
\draw [] (B') rectangle ([xshift=-1cm, yshift=1cm]B') node[pos=.5]{$+$};
\draw [] (D') rectangle ([xshift=1cm, yshift=-1cm]D') node[pos=.5]{$+$};
\draw [] (D') rectangle ([xshift=-1cm, yshift=1cm]D') node[pos=.5]{$+$};
\draw [] (F') rectangle ([xshift=1cm, yshift=-1cm]F') node[pos=.5]{$+$};
\draw [] (F') rectangle ([xshift=-1cm, yshift=1cm]F') node[pos=.5]{$+$};
\end{tikzpicture}
\times
\begin{tikzpicture}[scale=.4,baseline=(current bounding box.center)]
\coordinate (A)at(0,0);
\coordinate (B)at(0,10);
\coordinate (C)at(-5,10);
\coordinate (D)at(-5,7);
\coordinate (E)at(-10,7);
\coordinate (F)at(-10,3);
\coordinate (G)at(-15,3);
\coordinate (H)at(-15,0);
\coordinate (A')at(0,0);
\coordinate (B')at(0,6);
\coordinate (C')at(-3,6);
\coordinate (D')at(-3,3);
\coordinate (E')at(-8,3);
\coordinate (F')at(-8,0);
\draw (B')--(B)--(C)--(D)--(E)--(F)--(G)--(H)--(F')--(E')--(D')--(C')--(B') ;
\node at (-7,5) {$-Y$};
\draw [] (D) rectangle ([xshift=1cm, yshift=-1cm]D) node[pos=.5]{$-$};
\draw [] (D) rectangle ([xshift=-1cm, yshift=1cm]D) node[pos=.5]{$-$};
\draw [] (B) rectangle ([xshift=1cm, yshift=-1cm]B) node[pos=.5]{$-$};
\draw [] (B) rectangle ([xshift=-1cm, yshift=1cm]B) node[pos=.5]{$-$};
\draw [] (C) rectangle ([xshift=1cm, yshift=-1cm]C) node[pos=.5]{$+$};
\draw [] (C) rectangle ([xshift=-1cm, yshift=1cm]C) node[pos=.5]{$+$};
\draw [] (E) rectangle ([xshift=1cm, yshift=-1cm]E) node[pos=.5]{$+$};
\draw [] (E) rectangle ([xshift=-1cm, yshift=1cm]E) node[pos=.5]{$+$};
\draw [] (F) rectangle ([xshift=1cm, yshift=-1cm]F) node[pos=.5]{$-$};
\draw [] (F) rectangle ([xshift=-1cm, yshift=1cm]F) node[pos=.5]{$-$};
\draw [] (G) rectangle ([xshift=1cm, yshift=-1cm]G) node[pos=.5]{$+$};
\draw [] (G) rectangle ([xshift=-1cm, yshift=1cm]G) node[pos=.5]{$+$};
\draw [] (H) rectangle ([xshift=1cm, yshift=-1cm]H) node[pos=.5]{$-$};
\draw [] (H) rectangle ([xshift=-1cm, yshift=1cm]H) node[pos=.5]{$-$};
\draw [] (C') rectangle ([xshift=1cm, yshift=-1cm]C') node[pos=.5]{$-$};
\draw [] (C') rectangle ([xshift=-1cm, yshift=1cm]C') node[pos=.5]{$-$};
\draw [] (E') rectangle ([xshift=1cm, yshift=-1cm]E') node[pos=.5]{$-$};
\draw [] (E') rectangle ([xshift=-1cm, yshift=1cm]E') node[pos=.5]{$-$};
\draw [] (B') rectangle ([xshift=1cm, yshift=-1cm]B') node[pos=.5]{$+$};
\draw [] (B') rectangle ([xshift=-1cm, yshift=1cm]B') node[pos=.5]{$+$};
\draw [] (D') rectangle ([xshift=1cm, yshift=-1cm]D') node[pos=.5]{$+$};
\draw [] (D') rectangle ([xshift=-1cm, yshift=1cm]D') node[pos=.5]{$+$};
\draw [] (F') rectangle ([xshift=1cm, yshift=-1cm]F') node[pos=.5]{$+$};
\draw [] (F') rectangle ([xshift=-1cm, yshift=1cm]F') node[pos=.5]{$+$};
\end{tikzpicture}\,.\nonumber\\
\end{align}
We introduce the notations $A,B,\bar A, \bar B$ as in the previous section;
\begin{eqnarray}
\begin{tikzpicture}[scale=.4,baseline=(current bounding box.center)]
\coordinate (A)at(0,0);
\coordinate (B)at(0,-10);
\coordinate (C)at(5,-10);
\coordinate (D)at(5,-7);
\coordinate (E)at(10,-7);
\coordinate (F)at(10,-3);
\coordinate (G)at(15,-3);
\coordinate (H)at(15,0);
\draw (A)--(B)--(C)--(D)--(E)--(F)--(G)--(H)--(A) ;
\node at (7,-5) {$Y^{(1)}$};
\draw [] (A) rectangle ([xshift=-1cm, yshift=1cm]A);
\draw[dashed] (D)--([yshift=7cm]D);
\draw[dashed] (F)--([yshift=3cm]F);
\draw[bend right, distance=2.1cm] (A) to node [fill=white, inner sep=0.2pt,circle] {$s_{1}$} (B);
\draw[bend right, distance=2.1cm] ([yshift=7cm]D) to node [fill=white, inner sep=0.2pt,circle] {$s_{2}$} (D);
\draw[bend right, distance=2.1cm] ([yshift=3cm]F) to node [fill=white, inner sep=0.2pt,circle] {$s_{3}$} (F);
\draw[bend right] (A) to node [fill=white, inner sep=0.2pt,circle] {$r_{1}$} ([yshift=7cm]D);
\draw[bend left] (A) to node [fill=white, inner sep=0.2pt,circle] {$r_{2}$} ([yshift=3cm]F);
\draw[bend left] (A) to node [fill=white, inner sep=0.2pt,circle] {$r_{3}$} (H);
\draw [] (G) rectangle ([xshift=-1cm, yshift=1cm]G);
\draw [] (F) rectangle ([xshift=1cm, yshift=-1cm]F);
\draw[->,bend left,ultra thick] ([xshift=1.5cm,yshift=-1.5cm]F)node[right]{$A_{2}$}to([xshift=.5cm,yshift=-.5cm]F);
\draw[->,bend left,ultra thick] ([xshift=1.5cm,yshift=-1.5cm]G)node[right]{$B_{3}$}to([xshift=-.5cm,yshift=.5cm]G);
\draw[->,bend left,ultra thick] ([xshift=-2.5cm,yshift=2.5cm]A)node[left]{$b-b_{1}\varepsilon_{1}-b_{2}\varepsilon_{2}$}to([xshift=-.5cm,yshift=.5cm]A);
\end{tikzpicture}
\ \ \ \ 
\begin{tikzpicture}[scale=.4,baseline=(current bounding box.center)]
\coordinate (A')at(0,0);
\coordinate (B')at(0,-6);
\coordinate (C')at(3,-6);
\coordinate (D')at(3,-3);
\coordinate (E')at(8,-3);
\coordinate (F')at(8,0);
\draw (A')--(B')--(C')--(D')--(E')--(F')--(A');
\draw[dashed] (D')--([yshift=3cm]D');
\draw[bend right, distance=2.1cm] (A') to node [fill=white, inner sep=0.2pt,circle] {$\bar{s}_{1}$} (B');
\draw[bend left, distance=2.1cm] ([yshift=3cm]D') to node [fill=white, inner sep=0.2pt,circle] {$\bar{s}_{2}$} (D');
\draw[bend right] (A') to node [fill=white, inner sep=0.2pt,circle] {$\bar{r}_{1}$} ([yshift=3cm]D');
\draw[bend left] (A') to node [fill=white, inner sep=0.2pt,circle] {$\bar{r}_{2}$} (F');
\draw [] (D') rectangle ([xshift=1cm, yshift=-1cm]D');
\draw [] (E') rectangle ([xshift=-1cm, yshift=1cm]E');
\draw[->,bend left,ultra thick] ([xshift=1.5cm,yshift=-1.7cm]D')node[right]{$\bar{B}_{1}$}to([xshift=.5cm,yshift=-.5cm]D');
\draw[->,bend left,ultra thick] ([xshift=.5cm,yshift=-.7cm]E')node[right]{$\bar{A}_{2}$}to([xshift=-.5cm,yshift=.5cm]E');
\node at (1.5,-4) {$Y^{(2)}$};
\end{tikzpicture}
\end{eqnarray}
\ba
&&
\begin{cases}
A_{i}=(r_i+1)\varepsilon_1+(s_{i+1}+1)\varepsilon_2 +b-b_{1}\varepsilon_{1}-b_{2}\varepsilon_{2} \ (i=0,1,\cdots,f,r_0=0,s_{f+1}=0)\\
B_{i}=r_i\varepsilon_1+s_{i}\varepsilon_2+b-b_{1}\varepsilon_{1}-b_{2}\varepsilon_{2} \ (i=1,2,\cdots,f)
\end{cases}
\\
&&
\begin{cases}
\bar{B}_{i}=(\bar{r}_i+1)\varepsilon_1+(\bar{s}_{i+1}+1)\varepsilon_2+b-b_{1}\varepsilon_{1}-b_{2}\varepsilon_{2} \ (i=0,1,\cdots,\baf,\br_0=0,\bs_{\baf+1}=0)\\
\bar{A}_{i}=\bar{r}_i\varepsilon_1+\bar{s}_{i}\varepsilon_2+b-b_{1}\varepsilon_{1}-b_{2}\varepsilon_{2} \ (i=1,2,\cdots,\baf)
\end{cases}
\ea
Substituting $\hat{\phi}_I$'s in the explicit forms, the part of the integrand above becomes
\ba
& &\prod_{j=1}^n \frac{(\phi+\hat{\phi}_j)^2(\phi-\hat{\phi}_j)^2((\phi+\hat{\phi}_j)^2-\varepsilon^2)((\phi-\hat{\phi}_j)^2-\varepsilon^2)}{((\phi+\hat{\phi}_j)^2-\varepsilon_1^2)((\phi-\hat{\phi}_j)^2-\varepsilon_1^2)((\phi+\hat{\phi}_j)^2-\varepsilon_2^2)((\phi-\hat{\phi}_j)^2-\varepsilon_2^2)} \nn \\
& &= \frac{\prod_{i=1}^{f}(\phi^2-B_{i}^2)(\phi^2-(B_{i}+\varepsilon)^2)}{\prod_{i=0}^{f}(\phi^2-A_{i}^2)(\phi^2-(A_{i}-\varepsilon)^2)} \frac{\prod_{i=0}^{\bar{f}}(\phi^2-\bar{B}_{i}^2)(\phi^2-(\bar{B}_{i}-\varepsilon)^2)}{\prod_{i=1}^{\bar{f}}(\phi^2-\bar{A}_{i}^2)(\phi^2-(\bar{A}_{i}+\varepsilon)^2)}\nn.
\ea
Now we evaluate the residue which corresponds to adding 1-box into $Y$.
In terms of $Y_i$ ($i=1,2$), it corresponds to adding a box to $Y_1$
or deleting a box from $Y_2$.
 In the first case, the residue picks up the pole $A_{l}(l=0,\cdots,f)$ and the integrand above becomes
\ba
\frac{\tilde{Z}_{n+1,Y_+}}{n_{Y_+}}\left(\frac{\tilde{Z}_{n,Y}}{n_Y} \right)^{-1} &=&  \frac{-1}{2(n+1)}\frac{1}{\varepsilon_1\varepsilon_2} \nn \\
& & \frac{\prod_{i=1}^{f}(A_{l}^2-B_{i}^2)(A_{l}^2-(B_{i}+\varepsilon)^2)}{\prod_{i=0(\neq l)}^{f}(A_{l}^2-A_{i}^2)(A_{l}^2-(A_{i}-\varepsilon)^2)} \frac{\prod_{i=0}^{\bar{f}}(A_{l}^2-\bar{B}_{i}^2)(A_{l}^2-(\bar{B}_{i}-\varepsilon)^2)}{\prod_{i=1}^{\bar{f}}(A_{l}^2-\bar{A}_{i}^2)(A_{l}^2-(\bar{A}_{i}+\varepsilon)^2)}\label{rec1} \\
& &\times \frac{1}{2A_{l}(2A_{l}-\varepsilon)(A_{l}^2-b^2)}\,. \nn
\ea
In the second case, the residue picks up the pole {$-\bar{A}_{l}(l=1,\cdots,\bar{f})$ and the integrand above becomes
\ba
\frac{\tilde{Z}_{n+1,Y_+}}{n_{Y_+}} \left(\frac{\tilde{Z}_{n,Y}}{n_Y}\right)^{-1}&=&  \frac{-1}{2(n+1)} \frac{1}{\varepsilon_1\varepsilon_2} \nn \\
& & \frac{\prod_{i=1}^{f}(\bar{A}_{l}^2-B_{i}^2)(\bar{A}_{l}^2-(B_{i}+\varepsilon)^2)}{\prod_{i=0}^{f}(\bar{A}_{l}^2-A_{i}^2)(\bar{A}_{l}^2-(A_{i}-\varepsilon)^2)} \frac{\prod_{i=0}^{\bar{f}}(\bar{A}_{l}^2-\bar{B}_{i}^2)(\bar{A}_{l}^2-(\bar{B}_{i}-\varepsilon)^2)}{\prod_{i=1(\neq l)}^{\bar{f}}(\bar{A}_{l}^2-\bar{A}_{i}^2)(\bar{A}_{l}^2-(\bar{A}_{i}+\varepsilon)^2)}\label{rec2} \\
& &\times \frac{1}{2\bar{A}_{l}(2\bar{A}_{l}+\varepsilon)(\bar{A}_{l}^2-b^2)} \,.\nn
\ea
These are the recursion relations of the simplified models. 

We have a technical comment here. We have to be careful when we add
a box at a special location in $Y$.  For example,
when an extra box is added at
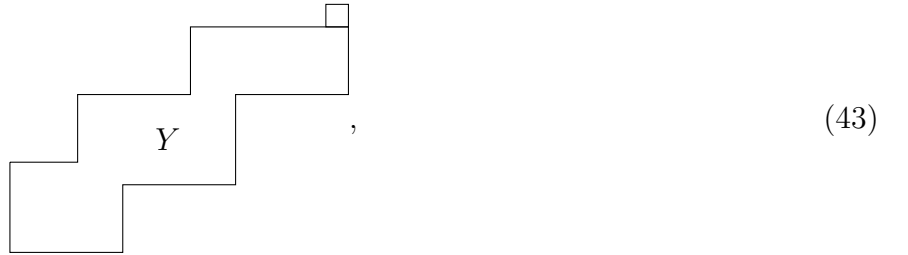
\ba
\begin{tikzpicture}[scale=.3,baseline=(current bounding box.center)]
\coordinate (A)at(0,0);
\coordinate (B)at(0,-10);
\coordinate (C)at(5,-10);
\coordinate (D)at(5,-7);
\coordinate (E)at(10,-7);
\coordinate (F)at(10,-3);
\coordinate (G)at(15,-3);
\coordinate (H)at(15,0);
\draw [] (H) rectangle ([xshift=-1cm,yshift=1cm]H);
\coordinate (A')at(0,0);
\coordinate (B')at(0,-6);
\coordinate (C')at(3,-6);
\coordinate (D')at(3,-3);
\coordinate (E')at(8,-3);
\coordinate (F')at(8,0);
\draw (B')--(B)--(C)--(D)--(E)--(F)--(G)--(H)--(F')--(E')--(D')--(C')--(B') ;
\node at (7,-5) {$Y$};
\end{tikzpicture}
,
\ea
the generalized Young diagram $Y = Y^{(1)}-Y^{(2)}$ is given as follows;
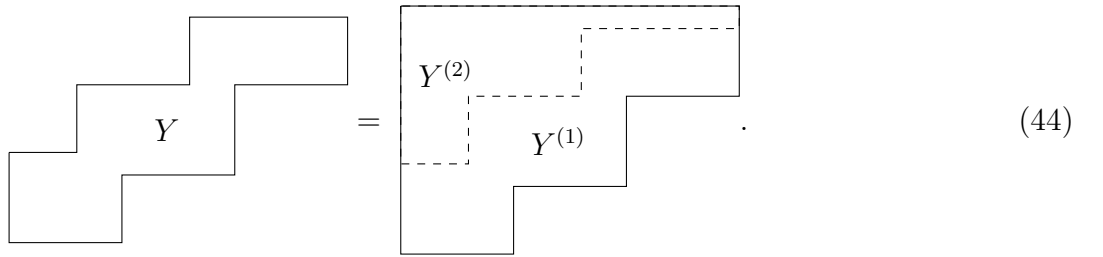
\ba
\begin{tikzpicture}[scale=.3,baseline=(current bounding box.center)]
\coordinate (A)at(0,0);
\coordinate (B)at(0,-10);
\coordinate (C)at(5,-10);
\coordinate (D)at(5,-7);
\coordinate (E)at(10,-7);
\coordinate (F)at(10,-3);
\coordinate (G)at(15,-3);
\coordinate (H)at(15,0);
\coordinate (A')at(0,0);
\coordinate (B')at(0,-6);
\coordinate (C')at(3,-6);
\coordinate (D')at(3,-3);
\coordinate (E')at(8,-3);
\coordinate (F')at(8,0);
\draw (B')--(B)--(C)--(D)--(E)--(F)--(G)--(H)--(F')--(E')--(D')--(C')--(B') ;
\node at (7,-5) {$Y$};
\end{tikzpicture}
=\
\begin{tikzpicture}[scale=.3,baseline=(current bounding box.center)]
\coordinate (A)at(0,1);
\coordinate (B)at(0,-10);
\coordinate (C)at(5,-10);
\coordinate (D)at(5,-7);
\coordinate (E)at(10,-7);
\coordinate (F)at(10,-3);
\coordinate (G)at(15,-3);
\coordinate (H)at(15,1);
\draw (A)--(B)--(C)--(D)--(E)--(F)--(G)--(H)--(A) ;
\node at (7,-5) {$Y^{(1)}$};
\coordinate (A')at(0,1);
\coordinate (B')at(0,-6);
\coordinate (C')at(3,-6);
\coordinate (D')at(3,-3);
\coordinate (E')at(8,-3);
\coordinate (F')at(8,0);
\coordinate(G')at(15,0);
\coordinate(H')at(15,1);
\draw[dashed, ] (A')--(B')--(C')--(D')--(E')--(F')--(G')--(H')--(A');
\node at (2,-2) {$Y^{(2)}$};
\end{tikzpicture}
.
\ea
In this case we have to add a row (instead of deleting a box) to $Y^{(2)}$.

Similarly, when an extra box is added at
\ba
\begin{tikzpicture}[scale=.3,baseline=(current bounding box.center)]
\coordinate (A)at(0,0);
\coordinate (B)at(0,-10);
\coordinate (C)at(5,-10);
\coordinate (D)at(5,-7);
\coordinate (E)at(10,-7);
\coordinate (F)at(10,-3);
\coordinate (G)at(15,-3);
\coordinate (H)at(15,0);
\draw [] (B) rectangle ([xshift=-1cm,yshift=1cm]B);
\coordinate (A')at(0,0);
\coordinate (B')at(0,-6);
\coordinate (C')at(3,-6);
\coordinate (D')at(3,-3);
\coordinate (E')at(8,-3);
\coordinate (F')at(8,0);
\draw (B')--(B)--(C)--(D)--(E)--(F)--(G)--(H)--(F')--(E')--(D')--(C')--(B') ;
\node at (7,-5) {$Y$};
\end{tikzpicture}
,
\ea
we have to add a column to $Y^{(2)}$;
\ba
\begin{tikzpicture}[scale=.3,baseline=(current bounding box.center)]
\coordinate (A)at(0,0);
\coordinate (B)at(0,-10);
\coordinate (C)at(5,-10);
\coordinate (D)at(5,-7);
\coordinate (E)at(10,-7);
\coordinate (F)at(10,-3);
\coordinate (G)at(15,-3);
\coordinate (H)at(15,0);
\coordinate (A')at(0,0);
\coordinate (B')at(0,-6);
\coordinate (C')at(3,-6);
\coordinate (D')at(3,-3);
\coordinate (E')at(8,-3);
\coordinate (F')at(8,0);
\draw (B')--(B)--(C)--(D)--(E)--(F)--(G)--(H)--(F')--(E')--(D')--(C')--(B') ;
\node at (7,-5) {$Y$};
\end{tikzpicture}
=\
\begin{tikzpicture}[scale=.3,baseline=(current bounding box.center)]
\coordinate (A)at(-1,0);
\coordinate (B)at(-1,-10);
\coordinate (C)at(5,-10);
\coordinate (D)at(5,-7);
\coordinate (E)at(10,-7);
\coordinate (F)at(10,-3);
\coordinate (G)at(15,-3);
\coordinate (H)at(15,0);
\draw (A)--(B)--(C)--(D)--(E)--(F)--(G)--(H)--(A) ;
\node at (7,-5) {$Y^{(1)}$};
\coordinate (A')at(-1,0);
\coordinate(B')at(-1,-10);
\coordinate(C')at(0,-10);
\coordinate (D')at(0,-6);
\coordinate (E')at(3,-6);
\coordinate (F')at(3,-3);
\coordinate (G')at(8,-3);
\coordinate (H')at(8,0);
\draw[dashed, ] (A')--(B')--(C')--(D')--(E')--(F')--(G')--(H')--(A');
\node at (2,-2) {$Y^{(2)}$};
\end{tikzpicture}
.
\ea
In the next section, we solve the relation and give the exact formula.
We have to be careful that the solution satisfies these exceptional
cases as well.

\section{Solution for the recursion relations}
In this section, we give a closed expression\footnote{Some years ago, there was a proposal of
	the instanton partition function
	for the BCD-type gauge theories \cite{Fucito:2004gi}
	in the self-dual background.
	It was based on the direct use of orbifold projection
	in the moduli space and the result takes
	very different form compared with ours.
	}
of the partition function associated with 
a generalized Young diagram
for the simplified integral with the anchor at $b$.
We prove that it satisfies the recursion relations 
(\ref{rec1},\ref{rec2}) derived in Sec.3.2.
We also made a cross-check explicitly for $n ( = [ \fr{k}{2} ] ) \leq 3$ cases.
\subsection{Proposed formula}
We will introduce some notations. We introduced the notions of the arm $a$ and leg $l$ 
of a Young diagram.  For the generalized Young diagram, it is more convenient to modify
the notation since we need to express the effect of the mirror:
\ba
u_p(s)&=&\phi_s+a_{Y^{(p)}}(s)\varepsilon_1=Y_j^{(p)}\varepsilon_1+j\varepsilon_2+(b-b_1\varepsilon_1-b_2\varepsilon_2)\\
v_p(s)&=&\phi_s+l_{Y^{(p)}}(s)\varepsilon_2=i\varepsilon_1+Y_i^{T(p)}\varepsilon_2+(b-b_1\varepsilon_1-b_2\varepsilon_2)
\ea
where $s=(i,j)$ is a position in a given Young diagram 
$Y^{(p)}$ $(p=1,2)$. 
When $s$ is in a generalized Young diagram $Y$ written as $Y^{(1)} - Y^{(2)}$, 
$u_p(s)$ and $v_p(s)$ represent the ``absolute coordinate" 
of the horizontal (vertical) border of $Y^{(p)}$ as follows:
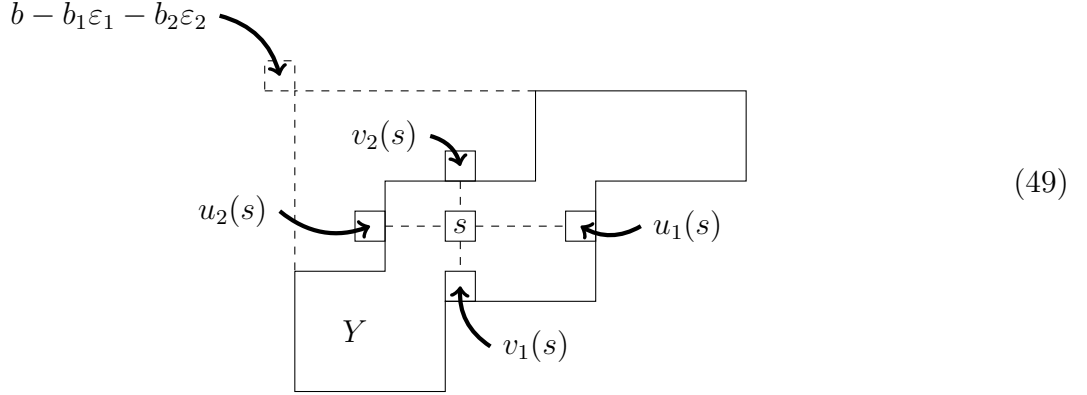
\ba
\begin{tikzpicture}[scale=.4,baseline=(current bounding box.center)]
\coordinate (A)at(0,0);
\coordinate (B)at(0,-10);
\coordinate (C)at(5,-10);
\coordinate (D)at(5,-7);
\coordinate (E)at(10,-7);
\coordinate (F)at(10,-3);
\coordinate (G)at(15,-3);
\coordinate (H)at(15,0);
\coordinate (A')at(0,0);
\coordinate (B')at(0,-6);
\coordinate (C')at(3,-6);
\coordinate (D')at(3,-3);
\coordinate (E')at(8,-3);
\coordinate (F')at(8,0);
\coordinate (X)at(5,-4);
\draw[dashed] (F')--(A')--(B');
\draw[dashed] ([xshift=-2cm,yshift=-.5cm]X)--([xshift=0cm,yshift=-.5cm]X);
\draw[dashed] ([xshift=1cm,yshift=-.5cm]X)--([xshift=4cm,yshift=-.5cm]X);
\draw[dashed] ([xshift=.5cm,yshift=1cm]X)--([xshift=.5cm,yshift=0cm]X);
\draw[dashed] ([xshift=.5cm,yshift=-1cm]X)--([xshift=.5cm,yshift=-2cm]X);
\draw (B')--(B)--(C)--(D)--(E)--(F)--(G)--(H)--(F')--(E')--(D')--(C')--(B') ;
\draw [dashed] (A) rectangle ([xshift=-1cm, yshift=1cm]A);
\draw[->,bend left,ultra thick] ([xshift=-2.5cm,yshift=2.5cm]A)node[left]{$b-b_{1}\varepsilon_{1}-b_{2}\varepsilon_{2}$}to([xshift=-.5cm,yshift=.5cm]A);
\draw (X) rectangle ([xshift=1cm, yshift=-1cm]X)node[pos=.5]{$s$};
\draw []([xshift=4cm]X) rectangle ([xshift=5cm, yshift=-1cm]X);
\draw []([yshift=-2cm]X) rectangle ([xshift=1cm, yshift=-3cm]X);
\draw []([yshift=2cm]X) rectangle ([xshift=1cm, yshift=1cm]X);
\draw []([xshift=-3cm]X) rectangle ([xshift=-2cm, yshift=-1cm]X);
\draw[->,bend left,ultra thick] ([xshift=-.5cm,yshift=2.5cm]X)node[left]{$v_{2}(s)$}to([xshift=.5cm,yshift=1.5cm]X);
\draw[->,bend right,ultra thick] ([xshift=-5.5cm,yshift=0cm]X)node[left]{$u_{2}(s)$}to([xshift=-2.5cm,yshift=-.5cm]X);
\draw[->,bend left,ultra thick] ([xshift=6.5cm,yshift=-.5cm]X)node[right]{$u_{1}(s)$}to([xshift=4.5cm,yshift=-.5cm]X);
\draw[->,bend left,ultra thick] ([xshift=1.5cm,yshift=-4.5cm]X)node[right]{$v_{1}(s)$}to([xshift=.5cm,yshift=-2.5cm]X);
\node at (2,-8) {$Y$};
\end{tikzpicture}
\ea
We claim that the partition function that solves the recursion relations is given as follows:
\ba
\tilde{Z}_{Y}(n)
=&& n_{Y} \fr{-1}{2^n n!} \label{l1}\\
&&\times \prod_{p,q=1,2}\left(
\prod_{s \in Y^{(p)}} \fr{u_p(s)+v_q(s)+\varepsilon}{u_p(s)-v_q(s)+\varepsilon_1}
\prod_{s \in Y^{(q)}} \fr{u_q(s)+v_p(s)}{-u_q(s)+v_p(s)+\varepsilon_2}
\right)^{(-1)^{p-q}}\label{l2}\\
&&\times \prod_{s\in Y^{(1)}}\fr{t(\phi_s+v_2(s))}{t(2\phi_s)t(-2\phi_s)}\times
\prod_{s\in Y^{(2)}}t(\phi_s+v_2^0(s) )\label{l3}\\
&&\times \prod_{s \in Y} \fr{((2\phi_s)^2-\varepsilon_1^2)((2\phi_s)^2-\varepsilon_2^2)}{\phi_s^2-b^2}\label{l4}
\ea
Here we use the following notations:
\begin{align}
\phi_s& = i \varepsilon_1 + j \varepsilon_2 + (b-b_1\varepsilon_1-b_2\varepsilon_2)\quad (\text{the absolute coordinate of the box $s$})\\
t(x)&=x(x+\varepsilon)(x-\varepsilon_1)(x+\varepsilon_2) \\
v_2^0(s)&=v_2(s)|_{Y_i^{T(2)}=0}=\phi_s-j\varepsilon_2
\end{align}

We note that the product in this formula should be treated carefully.
There are a few factors in the numerator and denominator that vanish. 
The numbers of zeros are identical in the numerator and denominator and should be
removed in the definition of formula.
For example, in the product $\prod_{s \in Y^{(1)}}(-u_1(s)+v_2(s)+\varepsilon_2)$  (\ref{l2}),
the factor associated with the upper right position of the diagram $Y$ vanishes.
Also the factor associated with the anchor of the diagram $Y$ vanish in the product (\ref{l4})).

We also note that the formula is not symmetric on $Y$ and $-Y$.
We do not know at this moment if such asymmetry is superficial and can be
removed by rewriting the formula.

In the rest of the section, we provide a proof that the formula satisfies the recursion formulae (\ref{rec1},\ref{rec2}).
We also cross-checked the formula by explicit computation for $n\leq 4$ by explicit computation.

\subsection{Proof of recursion relation}
We denote a diagram $Y$ with an extra box added by $Y_+$. 
We express $\tilde{Z}_{Y}(n)$ as $Z1(Y)\times Z2(Y)\times Z3(Y)\times Z4(Y)$, 
where $Z1$, (resp. $Z2$, $Z3$, $Z4$) is the factor 
in the line (\ref{l1}), (resp. (\ref{l2}), (\ref{l3}), (\ref{l4})), respectively.

We have to two possibility to add a box to $Y\ :\ $
adding a box to $Y=Y^{(1)}-Y^{(2)}$ means either adding a box to $Y^{(1)}$
or removing a box from $Y^{(2)}$.  
Since the proposed formula (\ref{l1}--\ref{l4}) is asymmetric with respect to $Y^{(1,2)}$ we need
separate analysis.

\paragraph{ (Case.1) Adding 1-box $\phi_{\hat{s}}(=A_l)$ into $Y$ along $Y_1$. ($Y^{(1)}_{\hat{m}}=r_l=\hat{n}-1$,$Y^{(1)T}_{\hat{n}}=s_{l+1}=\hat{m}-1$)}
The evaluation of the ratio of the factors
$Z1(Y_+)/Z1(Y)$, $Z3(Y_+)/Z3(Y)$, $Z4(Y_+)/Z4(Y)$
are straightforward: 
\ba
\fr{Z1(Y_+)}{Z1(Y)} &&= \fr{1}{2(n+1)},\\
\fr{Z3(Y_+)}{Z3(Y)} &&= \fr{t(\phi_{\hat{s}}+v_2(\hat{s}))}{t(2\phi_{\hat{s}})t(-2\phi_{\hat{s}})}=\fr{t((2\xi_1+2\hat{n},2\xi_2+\hat{m}+Y^{(2)T}_{\hat{n}}))}{(2A_l)^2((2A_l)^2-\varepsilon^2)((2A_l)^2-\varepsilon_1^2)((2A_l)^2-\varepsilon_2^2)},\\
\fr{Z4(Y_+)}{Z4(Y)} &&= \fr{((2A_l)^2-\varepsilon_1^2)((2A_l)^2-\varepsilon_2^2)}{A_l^2-b^2},
\ea
where $(x,y):=x\varepsilon_1+y\varepsilon_2 \ {\rm for} \ x,y \in \mathbb{C}$ and $b-b_1\varepsilon_1-b_2\varepsilon_2=:\xi_1\varepsilon_1+\xi_2\varepsilon_2$. 
On the other hand, lengthy calculation is needed to evaluate $Z2(Y_+)/Z2(Y)$.
After some analysis, we arrive at
\ba
\fr{Z2(Y_+)}{Z2(Y)}=&& \prod_{i=1}^{\hat{n}-1}\fr{
\tth(2\xi_1+\hat{n}+i,2\xi_2+\hat{m}+Y_i^{(1)T})
\tth(-\hat{n}+i,-\hat{m}+Y^{(1)T}_i)}{
\tth(2\xi_1+\hat{n}+i,2\xi_2+\hat{m}+Y^{(2)T}_i)
\tth(-\hat{n}+i,-\hat{m}+Y^{(2)T}_i)}
\label{tttt}
\\
&& \cdot \prod_{j=1}^{\hat{m}-1}
\ttv(2\xi_1+\hat{n}+Y^{(1)}_j, 2\xi_2+\hat{m}+j)
\ttv(-\hat{n}+Y^{(1)}_j, -\hat{m}+j)
\label{tt1}
\\
&& \cdot \prod_{j=1}^{Y^{(2)T}_{\hat{n}}}
\left(\ttv(2\xi_1+\hat{n}+Y^{(2)}_j, 2\xi_2+\hat{m}+j)
\ttv(-\hat{n}+Y^{(2)}_j, -\hat{m}+j)\right)^{-1}
\label{tt2}\\
&&\cdot (-1)\cdot \fr{2A_l(2A_l+\varepsilon)}{\varepsilon_1 \varepsilon_2}\\
&&\cdot \fr{(0,-\hat{m}+Y^{(2)T}_{\hat{n}}+1)(-1,-\hat{m}+Y^{(2)T}_{\hat{n}})}{
(2\xi_1+2\hat{n}+1,2\xi_2+\hat{m}+Y^{(2)T}_{\hat{n}}+1)(2\xi_1+2\hat{n},2\xi_2+\hat{m}+Y^{(2)T}_{\hat{n}})},
\ea
where
\ba
\tth(x,y):=\fr{(x,y)(x+1,y+1)}{(x-1,y)(x,y+1)}, \quad \ttv(x,y):=\fr{(x,y)(x+1,y+1)}{(x,y-1)(x+1,y)}.
\ea
Again we have cancellations between the factors in the numerator and the denominator.
For  that purpose the following formula is useful.  For arbitrary $x,y \in \mathbb{C},$
\ba
\prod_{i=1}^{\hat{n}-1}\tth(x+i,y+Y^{(1)T}_i)=&&\prod_{i=1}^{l}
\fr{(x+r_i,y+s_i)(x+r_i+1,y+s_i+1)}{(x+r_{i-1},y+s_i)(x+r_{i-1}+1,y+s_i+1)}
\label{f1}
\\
\prod_{j=1}^{\hat{m}-1} \ttv(x+Y^{(1)}_j,y+j)=&&\prod_{i=l+1}^{f}
\fr{(x+r_i, y+s_i)(x+r_i+1, y+s_i+1)}{(x+r_i,y+s_{i+1})(x+r_i+1,y+s_{i+1}+1)}.
\label{f2}
\ea
We also use the following identities: 
\ba
\prod_{i=1}^{\hat{n}-1} \left( \tth(x+i,y+Y^{(2)T}_i) \right)^{-1} =&& \prod_{i=1}^{\bar{l}} \fr{(x+\br_{i-1},y+\bs_{i})(x+\br_{i-1}+1,y+\bs_{i}+1)}{(x+\br_i,y+\bs_{i})(x+\br_i+1,y+\bs_i+1)} \\
&&\cdot \fr{(x+\br_{\bar{l}},y+\bs_{\bar{l}})(x+\br_{\bar{l}}+1,y+\bs_{\bar{l}}+1)}{(x+\hat{n}-1,y+Y^{(2)T}_{\hat{n}-1})(x+\hat{n},y+Y^{(2)T}_{\hat{n}-1}+1)}
\\
\prod_{j=1}^{Y^{(2)T}_{\hat{n}}} \left( \ttv(x+Y^{(2)}_j,y+j) \right)^{-1} =&& \prod_{i=\bar{l}'}^{\baf} \fr{(x+\br_i,y+\bs_{i+1})(x+\br_i+1,y+\bs_{i+1}+1)}{(x+\br_i,y+\bs_i)(x+\br_i+1,y+\bs_i+1)},
\ea
where $\bar{l}$, $\bar{l}'$ satisfy $\br_{\bar{l}-1}< \hat{n}-1 \leq \br_{\bar{l}}$, $\br_{\bar{l}'-1}< \hat{n} \leq \br_{\bar{l}'}$.
We arrive at:
\ba
\fr{Z2(Y_+)}{Z2(Y)}=&&\fr{\prod_{i=1}^f (A_{l}^2-B_i^2)(A_{l}^2-(B_i+\varepsilon)^2)}
{\prod_{i=0(\neq l)}^f(A_{l}^2-A_i^2)(A_{l}^2-(A_i-\varepsilon)^2)}\\
&&\cdot \fr{\prod_{i=0}^{\baf} (A_{l}^2-\bB_i^2)(A_{l}^2-(\bB_i-\varepsilon)^2)}
{\prod_{i=1}^{\baf} (A_{l}^2-\bA_i^2)(A_{l}^2-(\bA_i+\varepsilon)^2)}\\
&&\cdot (-1)\cdot \fr{2A_l(2A_l+\varepsilon)}{\varepsilon_1 \varepsilon_2} \cdot \fr{1}{t((2\xi_1+2\hat{n},2\xi_2+\hat{m}+Y^{(2)T}_{\hat{n}}))}.
\ea
Combining these results, we obtain the recursion relation (\ref{rec1}).

\paragraph{ (case.2) Removing a box $\phi_{\hat{s}}(=\bA_{\bar{l}})$ from $Y_2$} ($Y^{(2)}_{\hat{m}}=\br_{\bar{l}}=\hat{n}$,$Y^{(2)T}_{\hat{n}}=\bs_{\bar{l}}=\hat{m}$)

The variation of $Z1, Z3, Z4$ is straightforward:
\ba
\fr{Z1(Y_+)}{Z1(Y)} &&= \fr{1}{2(n+1)},\\
\fr{Z3(Y_+)}{Z3(Y)} &&= \prod_{j=1}^{Y_{\hat{n}}^{(1)T}} \fr{t((2\xi_1+2\hat{n},2\xi_2+\hat{m}-1+j))}{t((2\xi_1+2\hat{n},2\xi_2+\hat{m}+j))}\cdot \fr{1}{t((2\xi_1+2\hat{n},2\xi_2+\hat{m}))}\\
&&= \fr{1}{t((2\xi_1+2\hat{n},2\xi_2+\hat{m}+Y_{\hat{n}}^{(1)T}))},\\
\fr{Z4(Y_+)}{Z4(Y)} &&= \fr{((2A_l)^2-\varepsilon_1^2)((2A_l)^2-\varepsilon_2^2)}{\bA_{\bar{l}}^2-b^2}.
\ea
On the other hand, lengthy calculation gives:
\ba
\fr{Z2(Y_+)}{Z2(Y)}=&& \prod_{i=1}^{\hat{n}-1}\fr{
\tth(2\xi_1+\hat{n}+i,2\xi_2+\hat{m}+Y_i^{(1)T})
\tth(-\hat{n}+i,-\hat{m}+Y^{(1)T}_i)}{
\tth(2\xi_1+\hat{n}+i,2\xi_2+\hat{m}+Y^{(2)T}_i)
\tth(-\hat{n}+i,-\hat{m}+Y^{(2)T}_i)}
\\
&& \cdot \prod_{j=1}^{\hat{m}-1}
\left( \ttv(2\xi_1+\hat{n}+Y^{(2)}_j, 2\xi_2+\hat{m}+j)
\ttv(-\hat{n}+Y^{(2)}_j, -\hat{m}+j) \right)^{-1}
\\
&& \cdot \prod_{j=1}^{Y^{(1)T}_{\hat{n}}}
\ttv(2\xi_1+\hat{n}+Y^{(1)}_j, 2\xi_2+\hat{m}+j)
\ttv(-\hat{n}+Y^{(1)}_j, -\hat{m}+j)
\\
&&\cdot (-1)\cdot \fr{\varepsilon_1 \varepsilon_2}{2\bA_{\bar{l}}(2\bA_{\bar{l}}+\varepsilon)}\\
&&\cdot \fr{(2\xi_1+2\hat{n}+1,2\xi_2+\hat{m}+Y^{(1)T}_{\hat{n}}+1)(2\xi_1+2\hat{n},2\xi_2+\hat{m}+Y^{(1)T}_{\hat{n}})}{
(0,-\hat{m}+Y^{(1)T}_{\hat{n}}+1)(-1,-\hat{m}+Y^{(1)T}_{\hat{n}})}.
\ea
To simplify it, we need following formulae:
\ba
\prod_{i=1}^{\hat{n}-1} \left( \tth(x+i,y+Y^{(2)T}_i) \right)^{-1} =&& \prod_{i=1}^{\bk}
\fr{(x+\br_{i-1},y+\bs_i)(x+\br_{i-1}+1,y+\bs_i+1)}{(x+\br_i,y+\bs_i)(x+\br_i+1,y+\bs_i+1)}\\
&&\times \fr{(x+\br_{\bk},y+\bs_{\bk})(x+\br_{\bk}+1,y+\bs_{\bk}+1)}{(x+\br_{\bk}-1,y+\bs_{\bk})(x+\br_{\bk},y+\bs_{\bk}+1)}
\label{h1}
\\
\prod_{j=1}^{\hat{m}-1} \left( \ttv(x+Y^{(2)}_j,y+j) \right)^{-1} =&& \prod_{i=\bk}^{\baf}
\fr{(x+\br_{i}, y+\bs_{i+1})(x+\br_{i}+1, y+\bs_{i+1}+1)}{(x+\br_{i},y+\bs_{i})(x+\br_{i}+1,y+\bs_{i}+1)}\\
&&\times \fr{(x+\br_{\bk}, y+\bs_{\bk})(x+\br_{\bk}+1, y+\bs_{\bk}+1)}{(x+\br_{\bk},y+\bs_{\bk}-1)(x+\br_{\bk}+1,y+\bs_{\bk})}.
\label{h2}\\
\prod_{i=1}^{\hat{n}-1} \tth(x+i,y+Y^{(1)T}_i) =&& \prod_{i=1}^{l} \fr{(x+r_{i},y+s_{i})(x+r_{i}+1,y+s_{i}+1)}{(x+r_{i-1}-1,y+s_{i})(x+r_{i-1},y+s_{i}+1)} \\
&&\times \fr{(x+\hat{n}-1,y+Y^{(1)T}_{\hat{n}-1})(x+\hat{n},y+Y^{(1)T}_{\hat{n}-1}+1)}{(x+r_l,y+s_l)(x+r_l+1,y+s_l+1)}
\\
\prod_{j=1}^{Y^{(1)T}_{\hat{n}}} \ttv(x+Y^{(1)}_j,y+j) =&& \prod_{i=l'}^{f} \fr{(x+r_{i},y+s_{i})(x+r_{i}+1,y+s_{i}+1)}{(x+r_{i},y+s_{i+1}-1)(x+r_{i}+1,y+s_{i+1})},
\ea
where $l$, $l'$ satisfy $r_{l-1}< \hat{n}-1 \leq r_{l}$, $r_{l'-1}< \hat{n} \leq r_{l'}$. The variation of $Z2$ becomes:
\ba
\fr{Z2(Y_+)}{Z2(Y)}=&&\fr{\prod_{i=1}^f (\bA_{\bk}^2-B_i^2)(\bA_{\bk}^2-(B_i+\varepsilon)^2)}
{\prod_{i=0}^f(\bA_{\bk}^2-A_i^2)(\bA_{\bk}^2-(A_i-\varepsilon)^2)}\\
&&\cdot \fr{\prod_{i=0}^{\baf} (\bA_{\bk}^2-\bB_i^2)(\bA_{\bk}^2-(\bB_i-\varepsilon)^2)}
{\prod_{i=1(\neq \bk)}^{\baf} (\bA_{\bk}^2-\bA_i^2)(\bA_{\bk}^2-(\bA_i+\varepsilon)^2)}\\
&&\cdot (-1)\cdot \fr{ t((2\xi_1+2\hat{n},2\xi_2+\hat{m}+Y_{\hat{n}}^{(1)T}))}{\varepsilon_1 \varepsilon_2 2\bA_{\bk}(2\bA_{\bk}+\varepsilon)(4\bA_{\bk}^2-\varepsilon_1^2)(4\bA_{\bk}^2-\varepsilon_2^2)}.
\ea
Combining these results, we arrive at the recursion formula (\ref{rec2}).

\section{Discussion}
As we explained in the introduction and text, what we have derived is the contribution
of poles that may be interpreted as the contribution of D-brane.
There are many questions to be answered to understand the whole
picture of the problem.  One issue is how to derive the multiplicity factor $n_{\vec Y}$.
So far the general algorithm was not found for the integer factor.
Also we have not solved the contribution near the origin.
While the form of the recursion should be similar,
the location of the poles may overlap with their mirrors
and careful treatment is necessary to resolve it.
A related issue is the existence of extra configuration of poles which takes
the shape of cycles \cite{Hollands:2010xa}.  We suspect that
some extra inspirations will be necessary to solve these problems.

We considered the simplified integral in this paper because the original integral gives apparent double poles when we treat 8-instantons and higher cases, which seem to require us to take care of the appropriate integration cycle. Regarding this point, one may give correct results by counting JK-residues\cite{1993alg.geom..7001J,2004InMat.158..453S}, but it remains unclear whether we can count the multiplicity factor without looking at each pole.

Another puzzling feature of the problem is that the recursion relation 
does not directly related to the infinite dimensional symmetry
as in the $SU(N)$ case.  For $SU(N)$, the poles takes the form
of $N$ Young diagrams which can be identified with
the basis of the Hilbert space of $N$ bosons
and $W_N$ algebra.  For the generalized Young diagrams
and the contribution of the poles such obvious correspondence does
not seems to exist.  Also, the understanding of the multiplicity factor
from the viewpoint of field theory will be necessary.
We hope to come back to these issues in the feature publications.

Finally we comment on the relation with the nonlinear symmetry.
As we mentioned, for $SU(N)$ case, we have a direct relation
between the recursion relation and the action of generators
on the states.  For $BCD$ cases, the pole is labeled by the generalized
Young diagram.  So far, the precise correspondence
with the generalized Young diagram and the Fock space of 
some free fields is not obvious.  It should be also clarified
what would be the interpretation for the multiplicity factor
from the symmetry.
While double affince Hecke algebra has their analog in
arbitrary Lie group, it does not seem to have a direct implication
in these issues.  This is another issue that should be
studied in the future.

\section*{Acknowledgments}
We are obliged to Ivan Kostov, Kantaro Omori, Vincent Pasquier, Didina Serban, 
Hong Zhang and especially to Yuji Tachikawa
for the helpful discussion.  YM is partially  supported 
by Grants-in-Aid for Scientific Research (Kakenhi \#25400246)
and Fondation Math\'ematique Jacques Hadamard
for the support to stay at CEA Saclay where part of the study was made.
FO is supported by Advanced Leading Graduate Course for Photon Science grant.

\bibliographystyle{utphys}

\bibliography{reference}

\providecommand{\href}[2]{#2}\begingroup\raggedright\begin{thebibliography}{10}

\bibitem{Seiberg:1994rs}
N.~Seiberg and E.~Witten, ``{Electric - magnetic duality, monopole
  condensation, and confinement in N=2 supersymmetric Yang-Mills theory},''
  \href{http://dx.doi.org/10.1016/0550-3213(94)90124-4}{{\em Nucl.Phys.}
  {\bfseries B426} (1994) 19--52},
\href{http://arxiv.org/abs/hep-th/9407087}{{\ttfamily arXiv:hep-th/9407087
  [hep-th]}}.

\bibitem{Nekrasov:2003af}
N.~A. Nekrasov, ``{Seiberg-Witten prepotential from instanton counting},''
\href{http://arxiv.org/abs/hep-th/0306211}{{\ttfamily arXiv:hep-th/0306211
  [hep-th]}}.

\bibitem{Nekrasov2003}
N.~Nekrasov and A.~Okounkov, ``{Seiberg-Witten theory and random partitions},''
\href{http://arxiv.org/abs/hep-th/0306238}{{\ttfamily arXiv:hep-th/0306238
  [hep-th]}}.

\bibitem{Alday:2009aq}
L.~F. Alday, D.~Gaiotto, and Y.~Tachikawa, ``{Liouville Correlation Functions
  from Four-dimensional Gauge Theories},''
  \href{http://dx.doi.org/10.1007/s11005-010-0369-5}{{\em Lett.Math.Phys.}
  {\bfseries 91} (2010) 167--197},
\href{http://arxiv.org/abs/0906.3219}{{\ttfamily arXiv:0906.3219 [hep-th]}}.

\bibitem{Fateev:2009aw}
V.~Fateev and A.~Litvinov, ``{On AGT conjecture},''
  \href{http://dx.doi.org/10.1007/JHEP02(2010)014}{{\em JHEP} {\bfseries 1002}
  (2010) 014},
\href{http://arxiv.org/abs/0912.0504}{{\ttfamily arXiv:0912.0504 [hep-th]}}.

\bibitem{Alba:2010qc}
V.~A. Alba, V.~A. Fateev, A.~V. Litvinov, and G.~M. Tarnopolskiy, ``{On
  combinatorial expansion of the conformal blocks arising from AGT
  conjecture},'' \href{http://dx.doi.org/10.1007/s11005-011-0503-z}{{\em
  Lett.Math.Phys.} {\bfseries 98} (2011) 33--64},
\href{http://arxiv.org/abs/1012.1312}{{\ttfamily arXiv:1012.1312 [hep-th]}}.

\bibitem{schiffmann2013cherednik}
O.~Schiffmann and E.~Vasserot, ``Cherednik algebras, w-algebras and the
  equivariant cohomology of the moduli space of instantons on a 2,'' {\em
  Publications math{\'e}matiques de l'IH{\'E}S} {\bfseries 118} no.~1, (2013)
  213--342.

\bibitem{maulik2012quantum}
D.~Maulik and A.~Okounkov, ``Quantum groups and quantum cohomology,'' {\em
  arXiv preprint arXiv:1211.1287} (2012) .

\bibitem{Kanno:2013aha}
S.~Kanno, Y.~Matsuo, and H.~Zhang, ``{Extended Conformal Symmetry and Recursion
  Formulae for Nekrasov Partition Function},''
  \href{http://dx.doi.org/10.1007/JHEP08(2013)028}{{\em JHEP} {\bfseries 1308}
  (2013) 028},
\href{http://arxiv.org/abs/1306.1523}{{\ttfamily arXiv:1306.1523 [hep-th]}}.

\bibitem{Morozov2014}
A.~Morozov and A.~Smirnov, ``Towards the proof of agt relations with the help
  of the generalized jack polynomials,''
  \href{http://dx.doi.org/10.1007/s11005-014-0681-6}{{\em Lett.Math.Phys.}
  {\bfseries 104} no.~5, (2014) 585--612},
\href{http://arxiv.org/abs/1307.2576}{{\ttfamily arXiv:1307.2576 [hep-th]}}.

\bibitem{Matsuo:2014rba}
Y.~Matsuo, C.~Rim, and H.~Zhang, ``{Construction of Gaiotto states with
  fundamental multiplets through Degenerate DAHA},''
  \href{http://dx.doi.org/10.1007/JHEP09(2014)028}{{\em JHEP} {\bfseries 1409}
  (2014) 028},
\href{http://arxiv.org/abs/1405.3141}{{\ttfamily arXiv:1405.3141 [hep-th]}}.

\bibitem{Nekrasov:2004vw}
N.~Nekrasov and S.~Shadchin, ``{ABCD of instantons},''
  \href{http://dx.doi.org/10.1007/s00220-004-1189-1}{{\em Commun.Math.Phys.}
  {\bfseries 252} (2004) 359--391},
\href{http://arxiv.org/abs/hep-th/0404225}{{\ttfamily arXiv:hep-th/0404225
  [hep-th]}}.

\bibitem{Marino:2004cn}
M.~Marino and N.~Wyllard, ``{A Note on instanton counting for N=2 gauge
  theories with classical gauge groups},''
  \href{http://dx.doi.org/10.1088/1126-6708/2004/05/021}{{\em JHEP} {\bfseries
  0405} (2004) 021},
\href{http://arxiv.org/abs/hep-th/0404125}{{\ttfamily arXiv:hep-th/0404125
  [hep-th]}}.

\bibitem{Hollands:2010xa}
L.~Hollands, C.~A. Keller, and J.~Song, ``{From SO/Sp instantons to W-algebra
  blocks},'' \href{http://dx.doi.org/10.1007/JHEP03(2011)053}{{\em JHEP}
  {\bfseries 1103} (2011) 053},
\href{http://arxiv.org/abs/1012.4468}{{\ttfamily arXiv:1012.4468 [hep-th]}}.

\bibitem{Keller:2011ek}
C.~A. Keller, N.~Mekareeya, J.~Song, and Y.~Tachikawa, ``{The ABCDEFG of
  Instantons and W-algebras},''
  \href{http://dx.doi.org/10.1007/JHEP03(2012)045}{{\em JHEP} {\bfseries 1203}
  (2012) 045},
\href{http://arxiv.org/abs/1111.5624}{{\ttfamily arXiv:1111.5624 [hep-th]}}.

\bibitem{Fucito:2004gi}
F.~Fucito, J.~F. Morales, and R.~Poghossian, ``{Instantons on quivers and
  orientifolds},'' \href{http://dx.doi.org/10.1088/1126-6708/2004/10/037}{{\em
  JHEP} {\bfseries 0410} (2004) 037},
\href{http://arxiv.org/abs/hep-th/0408090}{{\ttfamily arXiv:hep-th/0408090
  [hep-th]}}.

\bibitem{Nakajima:2003pg}
H.~Nakajima and K.~Yoshioka, ``{Instanton counting on blowup. 1.},''
  \href{http://dx.doi.org/10.1007/s00222-005-0444-1}{{\em Invent.Math.}
  {\bfseries 162} (2005) 313--355},
\href{http://arxiv.org/abs/math/0306198}{{\ttfamily arXiv:math/0306198
  [math-ag]}}.

\bibitem{nakajimalectures}
H.~Nakajima, {\em Lectures on Hilbert Schemes of Points on Surfaces}.
\newblock University Lecture Series, vol. 18. American Mathematical Society,
  Providence, RI, 1999.

\bibitem{Shadchin:2005mx}
S.~Shadchin, ``{On certain aspects of string theory/gauge theory
  correspondence},''
\href{http://arxiv.org/abs/hep-th/0502180}{{\ttfamily arXiv:hep-th/0502180
  [hep-th]}}.

\bibitem{Atiyah:1978ri}
M.~Atiyah, N.~J. Hitchin, V.~Drinfeld, and Y.~Manin, ``{Construction of
  Instantons},''
\href{http://dx.doi.org/10.1016/0375-9601(78)90141-X}{{\em Phys.Lett.}
  {\bfseries A65} (1978) 185--187}.

\bibitem{Bruzzo:2002xf}
U.~Bruzzo, F.~Fucito, J.~F. Morales, and A.~Tanzini, ``{Multiinstanton calculus
  and equivariant cohomology},''
  \href{http://dx.doi.org/10.1088/1126-6708/2003/05/054}{{\em JHEP} {\bfseries
  0305} (2003) 054},
\href{http://arxiv.org/abs/hep-th/0211108}{{\ttfamily arXiv:hep-th/0211108
  [hep-th]}}.

\bibitem{Flume:2002az}
R.~Flume and R.~Poghossian, ``{An Algorithm for the microscopic evaluation of
  the coefficients of the Seiberg-Witten prepotential},''
  \href{http://dx.doi.org/10.1142/S0217751X03013685}{{\em Int.J.Mod.Phys.}
  {\bfseries A18} (2003) 2541},
\href{http://arxiv.org/abs/hep-th/0208176}{{\ttfamily arXiv:hep-th/0208176
  [hep-th]}}.

\bibitem{1993alg.geom..7001J}
L.~C. {Jeffrey} and F.~C. {Kirwan}, ``{Localization for nonabelian group
  actions},'' in {\em eprint arXiv:alg-geom/9307001}, p.~7001.
\newblock July, 1993.

\bibitem{2004InMat.158..453S}
A.~{Szenes} and M.~{Vergne}, ``{Toric reduction and a conjecture of Batyrev and
  Materov},'' \href{http://dx.doi.org/10.1007/s00222-004-0375-2}{{\em
  Inventiones Mathematicae} {\bfseries 158} (June, 2004) 453--495},
  \href{http://arxiv.org/abs/math/0306311}{{\ttfamily math/0306311}}.

\end{thebibliography}\endgroup
\end{document}